\documentclass[12pt,preprint]{aastex}

 \font\sevenrm=cmr7 scaled 1000

\makeatletter
\renewcommand{\fps@table}{t}

\makeatother

\renewcommand{\mag}{\mbox{$\;$mag}}

\shorttitle{Calibration of SNe Ia Hubble Diagram}
\shortauthors{Wang et al.}

\begin{document}

\title{Determination of the Hubble constant, the intrinsic
scatter of luminosities of Type Ia SNe, and evidence for
non-standard dust in other galaxies}

\author{Xiaofeng Wang\altaffilmark{1,2,3}, Lifan Wang\altaffilmark{4},
Reynald Pain\altaffilmark{3}, \\ Xu Zhou\altaffilmark{2}, Zongwei
Li\altaffilmark{5}} \altaffiltext{1}{Physics Department and
Tsinghua Center for Astrophysics (THCA), Tsinghua University,
Beijing, 100084, P R China; wang\_xf@mail.tsinghua.edu.cn}
\altaffiltext{2}{National Astronomical Observatories of China,
Chinese Academy of Sciences, Beijing 100012, P R China;
wxf@vega.bac.pku.edu.cn} \altaffiltext{3}{LPNHE, CNRS-IN2P3,
University of Paris VI\&VII, Paris, France} \altaffiltext{4}{E.O.
Lawrence Berkeley National Laboratory, 1 Cyclotron Rd., Berkeley,
CA94720, USA} \altaffiltext{5}{Department of Astronomy, Beijing
Normal University, 100875, Beijing, China}

\begin{abstract}
A sample of 109 type Ia supernovae (SNe Ia) with recession
velocity $\lesssim$ 30,000 km $s^{-1}$, is compiled from published
SNe Ia light curves to explore the expansion rate of the local
Universe. Based on the color parameter $\Delta C_{12}$ and the
decline rate $\Delta m_{15}$, we found that the average absorption
to reddening ratio for SN Ia host galaxies to be $R_{UBVI}$ =
4.37$\pm0.25$, 3.33$\pm0.11$, 2.30$\pm0.11$, 1.18$\pm0.11$, which
are systematically lower than the standard values in the Milky
Way. We investigated the correlations of the intrinsic luminosity
with light curve decline rate, color index, and supernova
environmental parameters. In particular, we found SNe Ia in E/S0
galaxies to be brighter close to the central region than those in
the outer region, which may suggest a possible metallicity effect
on SN luminosity. The dependence of SN luminosity on galactic
environment disappears after corrections for the extinction and
$\Delta C_{12}$. The Hubble diagrams constructed using 73 Hubble
flow SNe Ia yield a 1-$\sigma$ scatter of $\lesssim$0.12 mag in
$BVI$ bands and $\sim$0.16 mag in $U$ band. The luminosity
difference between normal SNe Ia and peculiar objects (including
SN 1991bg-like and 1991T-like events) has now been reduced to
within 0.15 mag via $\Delta C_{12}$ correction. We use the same
precepts to correct the nearby SNe Ia with Cepheid distances and
found that the fully corrected absolute magnitudes of SNe Ia are:
$M_{B}$ = $-$19.33$\pm$0.06$, M_{V}$ = $-$19.27$\pm$0.05.  We
deduced a value for the Hubble constant of H$_{0}$ = 72 $\pm$ 6
(total) km s$^{-1}$ Mpc$^{-1}$.

\end{abstract}


\section{Introduction}
Type Ia supernovae (SNe Ia) are probably the most precise distance
indicators known for measuring the extragalactic distances. Their
Hubble diagram, i.e., magnitude-redshift (m-z) relation, can be
used to trace the expansion history of the universe. The linear
portion of the Hubble diagram with absolute magnitude calibration
determines the Hubble constant (see a review in Branch 1998);
curvature in the diagram probes evolution of the expansion rate,
i.e., acceleration or deceleration, and consequently different
combinations of the cosmological parameters such as $\Omega_{m}$
and $\Omega_{\Lambda}$ (Riess et al. 1998; Perlmutter et al.
1999).

Various empirical methods have been developed to calibrate the
peak luminosities of SNe Ia. These include the template fitting or
$\Delta m_{15}$ method (Phillips 1993, Hamuy et al. 1996, Phillips
et al. 1999), the multi-color light curve method shape (MLCS)
method (Riess et al. 1996a; Jha 2002), and the stretch factor
method (Perlmutter et al. 1997; Goldhaber et al. 2001), the
"Bayesian Template Method" (BATM; Tonry et al. 2003), and the
recently proposed Spectral Adaptive Light Curve (SALT) method (Guy
et al. 2005). These methods are fundamentally identical by
utilizing the relationship between SN Ia light curve shape and
peak luminosity. The Color-Magnitude Intercept method (CMAGIC) of
Wang L et al. (2003, 2005) shows some variance, which replaces the
magnitude at maximum with the more uniform magnitude at a given
value of the color index. The above methods can yield distances to
SN Ia host galaxies with a relative precision approaching 8-11\%
which demonstrates the power of SNe Ia as cosmological lighthouses
for extragalactic distance scales. Wang et al. (2005) recently
proposed a method which is similar to the $\Delta m_{15}$ method
of Phillips et al. (1999), but instead of using the $B - V$ color
at maximum, we proposed to using the $B - V$ color at 12 days past
optical maximum as a more efficient calibration parameter.

By common practice so far, the spectroscopically peculiar SNe Ia
are usually excluded or given a lower weight in cosmological
studies. According to Li et al. (2001a), however, the total rate
of peculiar SNe Ia in a volume-limited search could be as high as
36\%. The rate of SN 1991T/1999aa-like objects is $\sim$20\% and
the rate of SN 1991bg-like objects is $\sim$16\%. Some of the
peculiar SNe Ia apparently deviate from the relation between light
curve shape and luminosity. The $\Delta C_{12}$ method provides a
better way to homogenize the normal SNe Ia and the
spectroscopically peculiar ones in a more unified and consistent
manner.

The color parameter $\Delta C_{12}\ = \ (B-V)_{12}$ gives tighter
empirical relations with SN Ia peak luminosities (Wang et al.
2005; W05). Here we apply this method to study the local Hubble
diagram by including different peculiar types of SNe Ia. Based on
the Hubble flow SNe Ia and nearby ones with Cepheid distances, we
deduce the value of the Hubble constant. In \S 2 we describe the
data selection for this study. In \S 3, we derive the host galaxy
reddening of SN Ia and estimate the ratios of extinction to
reddening for dust in SN host galaxies. In \S 4 we, examine the
luminosity dependence of SNe Ia on secondary parameters. In \S 5,
we present the Hubble diagram of SNe Ia and give the best
estimates of $H_{0}$. Conclusions and discussion are given in \S
6.

\section{The Data}
The major sources of SN Ia light curves  are: (a) the $BVI$ light
curves of 29 SNe Ia from the earlier Calan/Tololo SN survey (CTIO
sample; see Hamuy et al. 1996a,b,c), (b) the $BVRI$ light curves
of 22 SNe Ia collected by the CfA (CfA I sample; see Riess et al.
1999), (c) the $UBVRI$ light curves of another 44 SNe Ia from the
CfA SN monitoring campaign (CfA II sample; see Jha et al. 2006a),
and (d) the Las Campanas/CTIO observing campaign which covers
broad band $UBVRIJHK$ photometry (Krisciunas et al. 2001, 2003,
2004a,b).

Reindl et al. (2005; hereafter R05) compiled a sample of 124
nearby SNe Ia (z $\lesssim$ 0.1). Our sample is different in that
we include only SNe with CCD measurements, and those observed 8
days before maximum and with more than 5 photometric points. The
sample includes 109 SNe Ia, 88 of which are spectroscopically
normal (Branch et al. 1993), 13 are 91T/99aa-like (Phillips et al.
1992; Filippenko et al. 1992a; Li et al. 2001a), 6 are 91bg-like
(Filippenko et al. 1992b; Leibundgut et al. 1993). For light curve
fitting, we follow the template fitting procedure of Hamuy et al.
(1996b) but with six additional template SNe for better $\Delta
m_{15}$ sampling. Besides the SN templates (viz., 1991T, 1992bc,
1992al, 1992A, 1992bo, 1991bg) initially used by Hamuy et al.
(1996b), six additional SNe are included as templates: SNe 1999aa
($\Delta m_{15}$ = 0.82), 1999ee($\Delta m_{15}$ = 0.95),
1998aq-1998bu ($\Delta m_{15}$ = 1.05), 1996X-2002er($\Delta
m_{15}$ = 1.32), 2000dk ($\Delta m_{15}$ = 1.62), and 1999by
($\Delta m_{15}$ = 1.90). These light curve templates were also
used to derive the peak magnitudes in the $U$ band. Table 1 shows
the fitted parameters for this sample. They are tabulated in the
following manner:

\hspace{-0.35cm}
Column (1): The name of the each SN.\\
Column (2): The name of the corresponding host galaxy.\\
Column (3): The redshift in the reference frame of the cosmic
microwave background (CMB), using the procedure given at NASA
Extragalactic Database (NED; http://nedwww.ipac.caltech.edu/), or
for a self-consistent Virgo-centric infall vector of 220
km s$^{-1}$ taken from R05.\\
Column (4)-(7): The fitted peak magnitudes in $UBVI$ bands,
corrected for the Galactic reddening from Schlegel et al. (1998)
and the $K$ corrections from Nugent et al. (2002). The magnitude
errors in unit of 0.01 mag were a quadrature sum of the
uncertainties in the observed magnitudes, foreground reddening
(0.08 mag in $U$, 0.06 mag in $B$, 0.045 mag in $V$, and 0.03 mag
in $I$), as well as in the $K$ term (assumed to be 0.02 mag). \\
Column (8): The decline rate $\Delta m_{15}$, corrected for the small
reddening effect (Phillips et al. 1999).\\
Column (9): The $B - V$ color 12 days after $B$ maximum. It was
measured directly from the photometry or from the best-fit light
curve template when the observed color curves were too sparse to
be measured accurately. The $\Delta C_{12}$ value presented here
was already corrected for the Galactic reddening. \\
Column (10): The Galactic reddening following Schlegel et al. (1998).\\
Column (11): The reddening E$(B - V)$ in the host galaxy as
derived from the tail of the $B - V$ color curves.\\
Column (12): The reddening E$(B - V)$ in the host galaxy as
determined from the post maximum color $\Delta C_{12}$
(see next section and W05).\\
Column (13): The adopted reddening E$(B - V)_{host}$ in the host
galaxy which is the weighted average of Columns (11) and (12).\\
Column (14): The key reference sources of SN Ia photometry.\\

In Table 2, additional information is assembled for the
corresponding host galaxies of the SNe Ia listed in Table 1.
Columns (1)-(3) are self-explanatory. Column (4)-(5) list the
morphological and coded Hubble types of the host galaxies. Column
(6) gives the de-projected galactocentric distances of the SNe in
their respective host galaxies in units of the galaxy radius
r$_{25}$ (taken from the work of Xu \& Wang 2006 in preparation) .
Column (7)-(10) give the luminosity distances to the SNe Ia (see
discussion in $\S$ 5.3).

\section{The Extinction Correction}
The Galactic extinction are corrected by using the dust maps from
Schlegel et al. (1998). We need to know the intrinsic colors of
SNe to estimate the reddening by the dust in the host galaxies.
This unavoidably requires assumptions on the intrinsic properties
of SNe.

\subsection{The colors and reddening of SNe Ia}
It was shown by Lira (1995) that the intrinsic $B - V$ color
evolution of SNe Ia 30$-$90 days past $V$ maximum can be well
approximated by a simple linear relation: $(B - V)_{0} = 0.725 -
0.0118\times(t_{V} - 60)$, where $t_{V}$ is the time (in days)
from the $V$ maximum. By empirically assuming that the intrinsic
colors of SNe at late time are all identical and follow the above
relation, one can deduce the reddening from the offset of the
observed "tail" color of SN Ia from the above relation. The host
galaxy reddening E$(B - V)_{tail}$ estimated this way are reported
in column (11) of Table 1.

The "tail" color is difficult to measure accurately. In most
cases, we have to use the colors at maximum or shortly after
maximum, provided that their intrinsic behavior is understood. It
is difficult to define a reddening-free sample of SNe. Therefore
we have applied a color cut of E$(B - V)_{tail} \lesssim$ 0.06 mag
to construct a sub-sample of 36 SNe which are likely to suffer
little aborption. Figure~1 shows $\Delta m_{15}$ dependence of the
peak colors $U_{max} - B_{max}$, $B_{max} - V_{max}$, $V_{max} -
I_{max}$, and the post maximum color $\Delta C_{12}$. Half of
these SNe are located in the dust-poor earlier-type E/S0 galaxies.
80\% of the remaining half are located in the outskirts of their
spiral hosts.


As shown in Figure 1, all colors show similar "kinks" near $\Delta
m_{15}\sim1.65$. A simple linear relation failed to describe the
color-$\Delta m_{15}$ relation for these SNe Ia. A cubic spline is
employed to fit the data points in Fig.1. This results in a
dispersion of 0.07-0.08 mag for $BVI$ colors at maximum. The peak
$U - B$ color shows a wider range (from about $-$0.6 to 0.6) for
SNe with different $\Delta m_{15}$. The scatter in the $U - B$
color is also larger, e.g. $\sim$0.14 mag (see Jha et al. 2006a
for a similar argument by using the stretch factor).

Compared to the $UBVI$ colors at maximum light, the post-maximum
color parameter $\Delta C_{12}$ shows a much tighter dependence
on $\Delta m_{15}$, which is given as,
\begin{equation}
\Delta C_{12} = 0.33_{\pm0.01} + 0.32_{\pm0.04}(\Delta m_{15} -
1.1) - 0.56_{\pm0.21}(\Delta m_{15} - 1.1)^{2} + \\
2.16_{\pm0.22}(\Delta m_{15} - 1.1)^{3}, \sigma = 0.043
\end{equation}
The dispersion of $\sim$0.04 mag is comparable to the intrinsic
dispersion of the evolution of the "tail" color (e.g. Lira 1995,
Phillips et al. 1999), and is much less than that of the peak
color$-\Delta m_{15}$ relation. The residual distribution for the
fit to the color-$\Delta m_{15}$ correlation, as shown in Figure
2, reveals that the larger scatter of the peak colors is not due
to individual observations but tends to be an overall behavior.

Theoretically, the SN colors are related to the exact changes of
the optical depth in the ejecta. The peak colors of SNe Ia show
very rapid evolution and small measurement errors may result in
systematically incorrect reddening measurements (Leibundgut 2001).
For these reasons, we would prefer to use the post-maximum color
$\Delta C_{12}$ as an alternative reddening indicator. Remember
that Equation (1) is derived from SNe Ia with 0.8 $<$ $\Delta
m_{15}$ $<$ 2.0, and it may not apply to those SNe Ia with decline
rates beyond the above range (e.g., SN 2001ay, which has the
broadest light curve with $\Delta m_{15}$ = 0.69; see Howell \&
Nugent 2003). Deviation of the observed $\Delta C_{12}$ from the
curve shown in the bottom panel of Fig.1 gives an estimate of the
host galaxy reddening E$(B - V)_{12}$ which is listed in column
(12) of Table 1. The above determinations of the host galaxy
reddening are well consistent with those estimated by the color at
the nebular epoch, with an offset of $\sim$0.01 mag and a
dispersion of $\sim$0.04 mag.

The host galaxy reddening E$(B - V)_{host}$ was taken to be the
weighted average of E$(B - V)_{12}$ and E$(B - V)_{tail}$ (cf.
Phillips et al. 1999, Altavilla et al. 2004). In some cases, the
formal mean value turns out to be negative specifically, the E$(B
- V)_{host}$ values were assumed to be zero. This is equivalent to
adopting a Bayesian filter with a flat prior distribution for
positive E$(B - V)$ and zero for E$(B - V) < 0$ (Riess et al.
1996a). The adopted E$(B - V)_{host}$ values are listed in column
(13) of Table 1.

Figure 3 shows the distribution of the reddening E$(B - V)_{host}$
in their host galaxies (see $\S$ 4.1 as a function of the
normalized distance r$_{SN}$/r$_{25}$). The SNe further out are
found to have lower reddening values than those in the inner
regions. The mean E$(B - V)_{host}$ value for the SNe in E/S0
galaxies is found to be 0.07$\pm$0.02 mag. Excluding the largest
contributor SN 1986G in reddening, the mean reddening value still
remains at a non-negligible level of $\sim$0.05 mag.

We further compared the values of the host galaxy reddening
derived in this paper to the corresponding ones given in R05. As
shown in the lower panel of Fig.3, the mean value of R05 appears
to be lower than ours by 0.047$\pm$0.051 mag. This difference is
related to the assumption on the reddening-free SN sample and the
approach to estimate the host galaxy reddening. For example, we
define a low-reddening sample of SNe Ia using a color cut of E$(B
- V)_{tail}$ $\lesssim$ 0.06 mag, while R05 assumed that all SNe
in E/S0 galaxies are free from reddening. According to our
analysis, however, the SNe in the E/S0 galaxies may not
necessarily guarantee a reddening-free sample.

\subsection{Empirical reddening relations}
An understanding of the absorptions relies not only on the
availability of independent reddening indicators but also on the
knowledge of the dust properties. The determination of $R$ for the
reddening ratios from SNe Ia implies astonishingly different
optical properties for dust in distant galaxies. Earlier analysis
assuming that SNe Ia have a unique luminosity and color yielded
surprising smaller values of $R_{B}\sim$ 1.0$-$2.0 (e.g. Branch \&
Tammann 1992). Recent determinations of $R$ tend to give more
consistent values, e.g. $R_{B}$ = 3.55$\pm$ 0.30 (Riess et al.
1996b), 3.5$\pm$0.4 (Phillips et al. 1999), 3.5 (Altavilla et al.
2004), and 3.65$\pm$0.16 (Reindl et al. 2005). The slight
variations among these values are related to the ways of deriving
the host galaxy reddening for SNe Ia.

To examine the dust properties of distant galaxies using SNe Ia,
it is important to first remove the intrinsic dependence of SN Ia
luminosity on the light or color curve parameters. We have shown
in Fig.3 of W05 that the peak luminosities of SNe Ia with minimum
absorption can be well calibrated by the color parameter $\Delta
C_{12}$. The relation between $\Delta C_{12}$ and the absolute
magnitudes $M$ in $UBVI$ (which were obtained by assuming H$_{0}$
= 72 km s$^{-1}$ Mpc$^{-1}$), based on a sample of 33 Hubble flow
SNe Ia with E$(B - V)_{host}\lesssim 0.06$ mag as listed in Table
1 (including four 91T/99aa-like objects, four 91bg-like objects
and SN 2001ay), can be expressed in terms of linear relations:
\begin{equation}
M_{U} = -19.75_{\pm0.09} + 2.55_{\pm0.15}(\Delta C_{12} -
0.31), \\
N = 10, \sigma = 0.202 \\
\end{equation}
\begin{equation}
M_{B} = -19.30_{\pm0.02} + 1.93_{\pm0.07}(\Delta C_{12} -
0.31), \\
N = 33, \sigma = 0.108 \\
\end{equation}
\begin{equation}
M_{V} = -19.24_{\pm0.02} + 1.43_{\pm0.06}(\Delta C_{12} -
0.31), \\
N = 33, \sigma = 0.097 \\
\end{equation}
\begin{equation}
M_{I} = -18.97_{\pm0.02} + 1.01_{\pm0.06}(\Delta C_{12} -
0.31), \\
N = 30, \sigma = 0.101 \\
\end{equation}
The normalization to $\Delta C_{12}$ = 0.31 corresponds to the
color value of the fiducial supernova 1992al. The dispersion of
the $M - \Delta C_{12}$ relation for 24 normal SNe Ia with low
dust reddening decreases further down to 0.170 mag in the $U$
band, to 0.069 mag in $B$ band, to 0.068 mag in $V$ band, and to
0.069 mag in $I$ band, respectively. And the corresponding slopes
in $UBVI$ are 1.97$\pm$0.52, 1.72$\pm$0.11, 1.41$\pm$0.10, and
0.96$\pm$0.10, respectively, which are not inconsistent with those
shown in the above equations. This demonstrates the robustness of
the correlation between the post-maximum color $\Delta C_{12}$ and
the absolute magnitudes $M$, in particular in the $V$ and $I$
bands. The larger dispersion in the $U$ band may be caused by
small number statistics, the intrinsically larger luminosity
dispersion as suggested by Jha et al. (2006a), or both of them.

The $\Delta C_{12}$ procedure thus provides an independent and
more precise way to determine the average value of the reddening
ratio for distant galaxies hosting SNe Ia. The absolute magnitudes
M$_{UBVI}$ of all 109 SNe Ia, corrected for the Galactic
extinction and the intrinsic dependence on $\Delta C_{12}$ as
shown by Eqs.(2)-(5), are plotted against the values of their host
galaxy reddening in Figure 4.

SNe shown in Figure 4 with $v\lesssim$ 3,000 km s$^{-1}$
(represented by open squares) were not used for the fit of the
reddening vector due to the effect of the peculiar motions of
their host galaxies. Moreover, the two Hubble flow events SNe
1999ej and 2002cx were also not included in the fit. The former is
too faint for unknown reasons by about 0.6 mag for a normal
supernova. The latter is characterized by a 1991T-like pre-maximum
spectrum and a 1991bg-like luminosity, a normal $B - V$ color
evolution, and very low expansion velocities (Li et al. 2003),
which may be a new subclass of SNe Ia (Jha et al. 2006b). For the
linear fits in Figure 4, the resulting values of $R_{UBVI}$ are:
\begin{equation}
R_{U} = 4.36\pm0.25, R_{B} = 3.33\pm0.11, \\
R_{V} = 2.29\pm0.11, R_{I} = 1.18\pm0.11
\end{equation}
which are clearly smaller than the standard values for the Milky
Way. Our determinations of the $R$ values are slightly lower than
most of the recent determinations. From a sample of 62 Hubble flow
SNe Ia, Reindl et al. (2005) reported recently a value of
$R_{B}$=3.65$\pm$0.16. Although there were a lot of the same SNe
in our analysis, the host galaxy reddening E$(B - V)_{host}$ we
derived for these SNe was different (see \S 3.2). The smaller
reddening values obtained by Reindl et al. consequently led them
to derive the larger $R$ value. Note that they also derived their
E$(B - V)_{host}$ values from the peak $V - I$ colors, which
usually have larger scatter (e.g., 0.08-0.09 mag) and may be not
so reliable as a reddening indicator. Very recently, Wang L et al.
(2005) determined a lower value of R$_{B}$ = 2.59$\pm$0.24 from a
combining analysis of the maximum luminosity and the luminosity at
CMAGIC region. This value may parameterize well the effects both
of extinction by host galaxy and of intrinsic SN color dependence
on luminosity, but not necessarily the true $R$ value for distant
galaxies since they did not try to disentangle these two effects
in the analysis. Moreover, their analysis did not include two of
the most highly reddened SNe 1995E and 1999gd in the Hubble flow,
which may have significant impact on the final $R$ values.

The remarkably small scatter associated with the above $R_{UBVI}$
values indicates that the host galaxies for most of the SNe Ia
within z$\lesssim$0.1 do share fairly similar dust properties.
Besides the two Hubble flow SNe mentioned above, the highly
reddened SN 1999cl (not shown in Figure 4) is likely to be an
outlier, which deviates from the best fit line by $\sim$1.1 mag in
$B$ band. The disturbance of the peculiar motion cannot account
for the large deviation of SN 1999cl since its host galaxy NGC
4501 (M88) is the member galaxy of the Virgo cluster. Assuming an
average luminosity and a peculiar motion of $\lesssim$200 km
s$^{-1}$ for SN 1999cl, one finds R$_{B}$ = 2.40$\pm$0.37 for this
individual supernova. This indicates M88 may have a very
non-standard dust (see also Krisciunas et al. 2005 for an argument
of a low $R_{V}\sim$1.55 for SN 1999cl). SN 1997br also seems to
be an outlier, which was a spectroscopically peculiar, 91T-like
object (Li et al. 1999). Applying our $R$ values to correcting the
host galaxy extinction of SN 1997br would make it be
unrealistically bright, in particular in the $U$ band where it is
brighter than a normal SN Ia by $\sim$1.2 mag. This difference
probably suggests a very different dust property for the host
galaxy of SN 1997br, ESO 576-G40. However, there is still a large
uncertainty in the measurement of the recession velocity of ESO
576-G40 (see discussions in Li et al. 1999), which may also
contribute in part to the large deviation of SN 1997br as seen in
Fig.4. For the normal and low-reddening SN 2003du (Anupama et al.
2005), a peculiar motion of $\sim$800 km s$^{-1}$ relative to the
the CMB frame, is needed to explain its "outlying position" in
Fig.4.

\section{The Luminosity Properties of SNe Ia}
Estimates of the host galaxy reddening and determinations of the
empirical reddening relations, based on a larger SN Ia sample of
different sorts, would enable us to de-redden SNe Ia accordingly.
This will consequently help to reveal the nature of SNe Ia and to
determine the relevant parameters governing their optical
properties.

\subsection{The luminosity diversities}
In Figure 5, we plot the distribution of the extinction-corrected
absolute magnitudes in $UBVI$ with a bin size of 0.2 mag for all
109 SNe listed in Table 1.

As known in previous studies, different SNe Ia span a wider range
over their optical properties (e.g., see Leibundgut 2001 for a
review). It is clear from Fig.5 that both the maximum value and
the scatter of the luminosity are wavelength-dependent. Compared
with the luminosity of SNe Ia in the $I$ band, the luminosity in
the $U$ band is more luminous and the scatter is also larger. This
can be interpreted as that the SN Ia emission in the $U$ band may
be more sensitive to the variance of the progenitor properties,
such as the metallicity (Hoeflich et al. 1998; Lentz et al. 2000).

The spectroscopically peculiar, 91T-like and 99aa-like events lie
at the brighter end of the luminosity distribution of SNe Ia,
which are $\sim$0.3 mag brighter than the normal SNe Ia (Branch et
al. 1993). Moreover, these overluminous events seem to have
relatively uniform peak luminosity, with a small scatter of about
0.13 mag in the $BVI$ and 0.20 mag in the $U$ band. On the other
hand, the faintest SNe Ia are almost characterized by the
91bg-like events but for the most peculiar SN 2002cx. In the
contrary to the 91T/99aa-like SNe Ia, the 91bg-like events cover a
wider range of peak luminosities from $\sim$2 mag to $\sim$0.6 mag
in different wave bands. The luminosity distributions of SNe Ia in
$UBVI$ bands can be roughly fitted by Gaussians. The long tail at
the faint end indicates a continuous distribution of the peak
luminosities between the Branch-normal SNe Ia and the 91bg-like
events, possibly suggesting a common origin for their progenitors.

\subsection{The environmental effects on SN Ia luminosity}
Locations of the SNe Ia in their host galaxies may offer clues to
understand the origin of SN Ia optical diversities (e.g., Wang et
al. 1997), as the properties of the stellar populations [e.g.,
metallicity and age] may vary from galaxy to galaxy and vary with
the spatial positions within a galaxy. In Figure 6, we show the
radial distribution of the absolute magnitudes M$_{V}$, corrected
for the absorption in the Galaxy and the host galaxy as laid out
in $\S$3. The SN location (here we only considered the radial
position) is defined as the relative galactocentric distance
r$_{SN}$/r$_{25}$, where r$_{SN}$ is the de-projected radial
distance of the SNe Ia from the galactic center, and r$_{25}$ is
the de Vaucouleurs radius of that galaxy (e.g., de Vaucouleurs
1991).

To examine the age effect on SN Ia luminosity, we treat SNe in
late-types (spiral galaxies) and early-types (S0 and elliptical
galaxies) separately, where the stellar populations are assumed to
be different. We find that SN Ia luminosity is related to the
morphological types of the host galaxies. Brighter SNe Ia tend to
occur in spiral galaxies with younger stellar populations, while
most of the fainter events occur preferentially in the E/S0
galaxies with relatively older stars. This dichotomy was first
noticed by Hamuy et al. (1996c), and was later confirmed by Riess
et al. (1999) and Hamuy et al. (2000). This observational fact
argues for the age difference in the progenitors as the origin of
SN Ia optical diversities. Nevertheless, such an age effect alone
seems unlikely to account for the occurrence of the brighter
object SN 1998es (with $\Delta m_{15}$ = 0.87) in the lenticular
galaxy NGC 632 and the presence of the fainter event SN 1999by
(with $\Delta m_{15}$ = 1.90) in the Sb galaxy NGC 2841.

On the other hand, we did not find any radial variation for SN Ia
luminosity in spiral galaxies alone, which is consistent with
previous studies (Parodi et al. 2000; Ivanov et al. 2000). The SNe
in E/S0 galaxies (represented by open circles) at the first glance
do not show any significant correlation between the luminosity and
the radial position $r_{SN}/r_{25}$. Omitting arbitrarily those
peculiar SNe Ia, e.g. the 91bg-like events as labelled in Fig.5,
however, a trend with the galactocentric distances seems to emerge
for the others. A radial gradient of 0.30$\pm$0.12 was found for
normal SNe Ia in E/S0 galaxies, which is significant at a
confidence level of $\sim$2.5$\sigma$.

If the magnitude gradient observed for normal SNe Ia in E/S0
galaxies was true, then the radial metallicity variation in
elliptical galaxy (Henry \& Worthey 1999) may be responsible for
the diversities of SN Ia luminosities because the stellar
populations in these earlier galaxies are generally believed to be
coeval (Worthey 1994; Vazdekis et al. 1997). Theoretically, the
metallicity effect on the range of SN Ia luminosity may be
understood by the "strong wind" models proposed by Umeda et al.
(1999) who suggests that the lower-metallicity systems will tend
to have larger initial C/O masses and hence produce fainter SNe
Ia. The lack of magnitude gradient in spiral galaxies may be the
counteracting result from the effects both of age and metallicity
since the stellar populations in the inner regions of spiral
galaxies are, in general, both more metal rich and older in
relation to those in the outer regions (Henry \& Worthey 1999). In
order to further disentangle these two environmental aspects, one
may need to resort to the spectroscopic studies of SN Ia host
galaxies as suggested by Hamuy et al. (2000).

\subsection{Luminosity dependence of SNe Ia on the secondary parameters}
The correlations between SN Ia peak luminosities and the secondary
parameters were specifically investigated by Parodi et al. (2000)
and have recently been revisited by Reindl et al. (2005). Their
studies suggested that the decline rate $\Delta m_{15}$ and the
peak $B - V$ color (see also Tripp 1998; Tripp \& Branch 1999) are
the two key parameters for the homogenization of SN Ia
luminosities. To examine this relationship, the
absorption-corrected absolute magnitudes $M_{V}$ of 73 Hubble flow
SNe (see the definition in \S 5.1) are thus plotted against the
decline rate $\Delta m_{15}$, the peak  $B - V$ color, as well as
the post-maximum color $\Delta C_{12}$ in Figure 7. The 11 nearby
SNe Ia with cepheid distance calibrations from Table 5 are also
overlaid in the figure.

The dependence of SN Ia peak luminosity on the decline rate
$\Delta m_{15}$ becomes apparently nonlinear for the full SN
sample with different spectral types as shown in Fig.7a (the left
panel of Fig.7). Excluding SN 2001ay, however, a cubic spline
seems to give a perfect fit to the $M_{V} - \Delta m_{15}$
relation with a dispersion of $\sim$0.12 mag. The cubic polynomial
dependence and the small dispersion are expected if the color term
$\Delta C_{12}$ in Eq.(4) is substituted by the right side of
Eq.(1).

Fig.7b shows the $M_{V} - (B_{max} - V_{max})$ plot. Although the
peak color $B_{max} - V_{max}$ connects well with fainter SNe Ia
at the redder end, it is very loosely related to the peak
luminosity for most of the SNe Ia with $-0.1\lesssim B_{max} -
V_{max}\lesssim 0.1$ mag. The linear fit yields a larger
dispersion of $\sim$0.23 mag for the points in Fig.7b, which
indicates that the peak color alone cannot delineate the peak
luminosity of SNe Ia. Combining both $\Delta m_{15}$ and $B_{max}
- V_{max}$ in the homogenization can reduce the luminosity
dispersion to $\sim$0.16 mag, which is consistent with what
obtained by R05.

The correlation of the extinction-corrected absolute magnitudes
with $\Delta C_{12}$ is illustrated in Fig 7c. As it can be seen,
the tight and linear $M_{V}- \Delta C_{12}$ relation found for the
reddening-free SNe still holds well for the larger SN sample. The
rms scatter yields along the best fit line is $\lesssim$0.12 mag
in $BVI$, which corresponds to a distance uncertainty of 5-6\%. As
the host galaxy reddening E$(B - V)_{host}$ was also partially
derived from the observed color $\Delta C_{12}$, it is necessary
to examine the possible interplay of these two variables. The
corresponding regression with these two variables takes the form
of
\begin{equation}
M_{UBVI} = b_{UBVI}(\Delta C_{12} - E(B - V)_{host} - 0.31)+
R_{UBVI}E(B - V)_{host} + M^{0}_{UBVI}
\end{equation}
leading further to
\begin{equation}
M_{UBVI} = b_{UBVI}(\Delta C_{12} - 0.31)+ (R_{UBVI} -
b_{UBVI})E(B - V)_{host} + M^{0}_{UBVI}
\end{equation}
where the constant term, $M^{0}_{UBVI}$, is the mean absolute
magnitude reduced to $\Delta C_{12}$ = 0.31. Least-squares
solutions for all 73 Hubble flow SNe in Fig.7 give the values of
$b_{UBVI}$, $R_{UBVI}$, and $M^{0}_{UBVI}$ as shown in Table 3.
The improved $R_{UBVI}$ values are very close to the provisional
values determined from Fig.4, and the slopes of the correlation
$b_{UBVI}$ agree quite well with those shown in Eqs.(2)-(5). This
suggests that the two variables E$(B - V)_{host}$ and $\Delta
C_{12}$ are quite independent of each other. Note that the slope
of the correlation and the reddening ratio show very similar
values in the $I$ band. In other words, we can homogenize well the
$I-$band luminosity of SNe Ia by the single parameter $\Delta
C_{12}$ without knowing the host galaxy reddening (because the
second term of the right side of Eq.(8) approximately drops out).

The small dispersion given by the $M - \Delta C_{12}$ relation
(see Table 3 and Fig.7) leaves little room for the dependence of
SN Ia luminosity on other secondary parameters. To further
strengthen our case, we examine the remaining dependence of the
$M_{V} - \Delta C_{12}$ relation fits on other secondary
parameters (see Figure 8). After homogenization as to $\Delta
C_{12}$ according to Eq.(7), the magnitude residuals
$\delta$M$_{V}$ from the best fit in Fig.7c did not show any
significant dependence on the decline rate $\Delta m_{15}$, or on
the peak $B - V$ color. As it can be seen from Fig.8, the
luminosity gradient with the galactocentric distances
r$_{SN}$/r$_{25}$ found for SNe Ia in E/S0 galaxies also becomes
marginally important with a confidence level of $\lesssim
1\sigma$. This shows that the environmental effects (such as the
metallicity and /or the age of the progenitors) on SN peak
luminosity can be removed or substantially reduced via $\Delta
C_{12}$ calibration.

The $M - \Delta C_{12}$ relation is re-examined and confirmed for
a larger SN sample, which allows us to calibrate SNe Ia using
$\Delta C_{12}$ with more confidence. It should be emphasized that
this is still not a universal relation, as there are still
extremely "anomalous" or rare events, such as SN 2002cx which
cannot be calibrated by any known methods.

\section{The distance scales}
\subsection{The Hubble diagram of SNe Ia}
An effective way of assessing the quality of SNe Ia as distance
indicators is to plot them in the Hubble diagram. The two
quantities entering the Hubble diagram $-$ magnitude $m$ and
redshift $z$$-$ are direct tracers for the expansion history of
the Universe. Our data sample consists of 109 well-observed SNe
Ia; nevertheless, not all of them are suitable for constructing
the Hubble diagram and exploring the expansion rate of the
Universe. We have intentionally excluded those SNe Ia in galaxies
with $cz$ $\lesssim$ 3,000 km s$^{-1}$ to reduce the uncertainties
from the peculiar motions. In addition, we further excluded SN
1996ab with $cz$ = 37370 km s$^{-1}$, where the cosmological
effect on the luminosity distance becomes important (Jha et al.
1999). After these selections, the remaining sample consists of 73
SNe Ia in the Hubble flow.

At distances of z$\lesssim$0.1, the dimming of a standard candle
as function of redshift $z$ can be simply described by
\begin{equation}
m = 5logcz + \alpha,
\end{equation}
where $cz$ is the recession velocity corrected to the CMB frame,
and $\alpha$ is the "intercept of the ridge line" given by
\begin{equation}
\alpha = 5log H_{0} -  M - 25.
\end{equation}
It follows immediately that
\begin{equation} H_{0} = 10^{0.2(M\
+\  \alpha\  +\  25)}.
\end{equation}
By Equation (11), determination of the Hubble constant $H_{0}$
from SNe requires their absolute magnitudes and the intercept of
their Hubble lines. The absolute magnitudes are obtained by
observations of nearby SNe Ia with Cepheid distances (see \S 5.2),
while the intercept $\alpha$ can be determined by observations of
the Hubble flow ones.

In Figure 9, we present the $U-$, $B-$, $V-$, and $I-$band Hubble
diagrams (or $m - z$ relation) for the 73 Hubble flow SNe Ia. The
apparent magnitudes are corrected for the $\Delta C_{12}$
dependence and the host galaxy absorption according to Eq.(7). The
small scatter in the Hubble diagram results in precise
determination of the $\alpha$ value in Eq.(9), which are
$-$4.18$\pm$0.03 in $U$, $-$3.64$\pm$0.01 in $B$, $-$3.56$\pm$0.01
in $V$, and $-$3.25$\pm$0.01 in $I$, respectively. To test the
robustness of the global fit, we examined the $m - z$ relation
using different subsets of SN sample in the Hubble flow. This
includes the recession velocity, the host galaxy reddening, and
the spectral types of SNe Ia. The best fit results are summarized
in Table 4. Note that the number shown in column (2) of Table 2
refers to the SN number with available data in the $B$ and $V$
bands.

The Hubble lines defined by SNe Ia with different spectral types
(e.g., normal, 91T/99aa-like, and 91bg-like) may have remarkable
offsets in the zeropoints, due to the larger difference between
their intrinsic luminosities (see Fig. 5). As a test of the
spectral dependence, we examined the Hubble diagrams constructed
by normal, 91T/99aa-like, and 91bg-like SNe Ia, respectively. It
turns out that the $\alpha$ values derived from the peculiar SNe
Ia are consistent with the global fit within $\pm$0.15 mag (see
Table 4). Including a fraction of $\sim$20\% peculiar events in
the full SN sample appears not to affect the Hubble line. This is
demonstrated by the statistically insignificant variations in the
$\alpha$ values between the full sample and the sample of 58
normal ones.

As the normal SNe Ia look more uniform, with a small scatter of
about 0.10 mag in $UBVI$ bands, they are appropriately used to
probe the possible variation in the expansion rate. Zehavi et al.
(1998) suggested a dynamic glitch in the Hubble diagram of SNe Ia
at cz = 7,000 km s$^{-1}$, across which the expansion rate of the
local Universe may change by a few percent. To examine this effect
and the possible variation in $\alpha$, we divided the sample of
58 normal SNe Ia into two subsample using the velocity cut at
7,000 km s$^{-1}$. The variations of the $\alpha$ values between
the more distant SNe Ia and the nearby ones are found to be around
0.03 mag in $BVI$ bands, which would lead to a discrepancy in the
expansion rate less than 2\%. The larger difference shown in the
$U$ band is more likely due to a statistical fluke since there are
only 7 SNe Ia with $U$-band photometry beyond v = 7,000 km
s$^{-1}$. However, more Hubble flow SN Ia sample and various
analysis techniques are still needed to further diagnose the
possible discontinuity of the expansion rate at the local
Universe.

To explore the effect of the absorption corrections on the Hubble
line, we considered the cases with restrictions on the maximum
value of E$(B - V)_{host}$ in the fitting. Excluding the highly
reddened SNe [i.e., E$(B - V)_{host}$ $>$ 0.15 mag], or even the
SNe with E$(B - V)_{host}$ $>$ 0.06 mag from the Hubble diagram,
the change in the $\alpha$ value is $\lesssim$0.02 mag in each
waveband. This argues in favor of the self-consistent
determinations of the host galaxy reddening and the unconventional
$R_{UBVI}$ values. Assuming the scatters shown by the fit to SNe
Ia with minimum absorption [e.g., E$(B - V)_{host}\leq 0.06$ mag]
as the intrinsic ones, we can estimate the uncertainties caused by
the absorption corrections. By comparing them with those values
shown in the fit of all normal SNe Ia, we place a limit on the
errors of the absorption corrections as $\sim$0.08, 0.07, and 0.06
mag in $B$, $V$, and $I$ bands, respectively. This is slightly
smaller than the corresponding values derived from the errors of
the host galaxy reddening given in Table 1. The small number of
statistic sample prevented us from giving a reasonable value in
the $U$ band.

It is satisfactory that the Hubble flow SNe Ia of different
subsets yield the same $\alpha$ values in each waveband within the
errors, showing the robustness of the $\Delta C_{12}$ method. In
particular, the dispersion around the Hubble line is impressively
small for each subset of SN sample. For example, the dispersion
for the sample of 58 normal SNe Ia is only 0.09-0.11 mag in full
optical bands, corresponding to a $\sim$5\% relative distance
uncertainty per supernova\footnote{Considering the typical
uncertainties of the redshift due to peculiar velocity (e.g. 300
km s$^{-1}$ adopted in our analysis), the actual uncertainty
intrinsic to our $\Delta C_{12}$ distance calibration is less than
4\%, which can be fully interpreted as the photometric errors.}.
For comparison, the MLCS method yields $\sigma_{B}$ = 0.22 mag for
their gold sample of 67 SNe Ia (Riess et al. 2004), whereas the
BATM calibration yields $\sigma_{B} = 0.21$ mag (BATM; Tonry et
al. 2003). The two-parameter method (which involves both $\Delta
m_{15}$ and peak $B - V$ color corrections) revisited recently by
Reindl et al. (2005) gives a dispersion of 0.16 mag for 63 SNe Ia.
In an approximately parallel analysis, Wang L et al. (2005)
obtained an rms scatter of $\sigma_{B} \approx$ 0.14 mag for 33
selected SN sample. The SALT method (Guy et al. 2005) gives
$\sigma_{B}$ = 0.16 mag for a sample of 23 SNe Ia. In reference to
above calibration methods, the $\Delta C_{12}$ method improves
remarkably the distance accuracy from SNe Ia. This will invoke
hopefully more precise determinations of the Hubble constant
H$_{0}$ and other cosmological parameters. The application of the
$\Delta C_{12}$ method to the Supernova Legacy Survey data (Astier
et al. 2006) will be presented in a forthcoming paper.

\subsection{Calibrations of the absolute magnitudes of SNe Ia}
In order to infer the Hubble constant $H_{0}$, it is necessary to
tie our $\Delta C_{12}$ distances to the SNe Ia with absolute
magnitude calibrations. Cepheid variables, presently, through
their Period-Luminosity ($P-L$) relation, are the fundamental
primary distance indicators. We can thus establish the absolute
magnitudes of SNe Ia through the distances measured from the
Cepheid variables in their host galaxies.

Thanks primarily to the valuable contribution of the
Saha-Tammann-Sandage SN Ia HST Calibration Project (hereafter
STS), the number of SNe Ia with Cepheid distances to their host
galaxies increased to thirteen so far. The STS consortium provided
the Cepheid distances to SNe 1895B (NGC 5253), 1937C (IC 4182),
1972E (NGC 5253), 1981B (NGC 4536), 1960F (NGC 4496A), 1990N (NGC
4639), 1989B (NGC 3627), 1998aq (NGC 3982) and 1991T (NGC 4527)
(Saha et al. 1996, 1997, 1999, 2001a, 2001b). The Cepheid
distances to SN 1974G in NGC 4414, SN 1998bu in NGC 3368, and SN
1999by in NGC 2841 was obtained by Turner (1998), Tanvir et al.
(1999), and Macri et al. (2001), respectively. Using the Advanced
Camera for Surveys (ACS) on HST, Riess et al. (2005) recently
measured the Cepheid distance to the more distant SN 1994ae in NGC
3370. These SNe Ia with Cepheid distances will provide the
calibration for the absolute magnitudes of SNe Ia.

Nevertheless, analysis of the preceding Cepheid data does not
reach a consistent result on the absolute calibration for SN Ia
peak luminosity (see also discussions by Riess et al. 2005).
Compared with the results of STS, the distances reanalyzed by HST
key project (hereafter KP) are on average shorter by 0.2$-$0.3 mag
(Gibson et al. 2000; Freedman et al. 2001). This discrepancy may
arise primarily from the sample selection of the Cepheid variables
and the choice of the $P-L$ relation (i.e., Saha et al. 2001a).
The STS consortium used the earlier $P-L$ relation that was
established by Madore \& Freedman (1991; hereafter MF91), while
the KP group employed a new $P-L$ relation which is based on the
Cepheid data of the Optical Gravitational Lensing Experiment
(OGLE) survey (Udalski 1999). The OGLE $P-L$ relation, with a
flatter color-period relation of $(V - I)$ $\propto$ 0.2log$P$,
yields a shorter Cepheid distance by about 8\% than that derived
from the earlier MF91 relation. We preferred the OGLE-based $P-L$
relation and the resulting Cepheid distances (i.e., the KP
distances; Freedman et al. 2001) in this paper, due to the
well-sampled light curves and the accurate photometry of the OGLE
Cepheids (Udalski et al. 1999; Sebo et al. 2002).

The Cepheid $P-L$ relation derived from LMC was traditionally
considered to be universal for all other galaxies. However, there
is increasing evidence for a significant dependence of the Cepheid
properties on the metallicity of the host stellar populations. An
empirical test of this dependence by Kennicutt et al. (1998)
yielded $\Delta \mu_{0}$ /$\Delta$[O/H] = $-$0.24$\pm$0.16 mag
dex$^{-1}$. Such a dependence is in agreement with the theoretical
predictions by Fiorentino et al. (2002). Using a larger sample of
17 Cepheid hosts with independent distances, Sakai et al. (2004)
obtained a more robust empirical correction relation $\Delta
\mu_{0}$ /$\Delta$[O/H] = $-$0.24$\pm$0.05 mag dex$^{-1}$ with
reduced errors, which is adopted here as metallicity correction
for the Cepheid distances.

Table 5 contains a list of partial parameters for 11 nearby SNe Ia
with Cepheid distances (see Table 1 for other parameters in more
details). The two historical SNe 1895 and 1960F were not included
because their $V$ light curves are unavailable to infer the
$\Delta C_{12}$ value, and hence the host galaxy reddening. The KP
distance moduli of these calibrators, with and without the
metallicity corrections, are given in columns (3) and (4). The
apparent $V$ magnitudes (corrected for the Galactic absorption),
the color parameter $\Delta C_{12}$, and the host galaxy reddening
are listed in columns (5)-(7). The fiducial magnitudes
m$^{0}_{V}$, corrected for absorption and $\Delta C_{12}$ using
Eq.(8), are given in column (8). The final absolute magnitudes of
the calibrators, M$^{0}_{U}$, M$^{0}_{B}$, M$^{0}_{V}$, and
M$^{0}_{I}$ contained in columns (9)-(12), respectively, are
derived by subtracting the distance moduli $\mu_{KP}$(Z) from the
corresponding apparent magnitudes m$^{0}_{UBVI}$ (here we only
listed in Table 5 the case for apparent $V$ magnitudes for the
space limit). The uncertainties associated with the absolute
magnitudes for each calibrator were obtained by combining in
quadrature the quoted errors $\delta$m, $\delta$$\Delta C_{12}$,
$\delta$E$(B - V)_{host}$, and $\delta$$\mu_{KP}(Z)$.

The weighted average of the peak absolute magnitudes in $UBVI$ for
the 11 nearby calibrators, reduced to $\Delta C_{12}$ = 0.31 using
Eq.(8) and the coefficients listed in Table 3, are given as
\begin{equation}
M_{U} = - 19.89\pm0.08 \mag, \ \sigma_{U} = 0.14 \mag
\end{equation}
\begin{equation}
M_{B} = - 19.33\pm0.06 \mag, \ \sigma_{B} = 0.11 \mag
\end{equation}
\begin{equation}
M_{V} = - 19.27\pm0.05 \mag, \ \sigma_{V} = 0.11 \mag
\end{equation}
\begin{equation}
M_{I} = - 18.97\pm0.04 \mag, \ \sigma_{I} = 0.08 \mag
\end{equation}
Excluding the spectroscopically peculiar SNe 1991T and 1999by, or
further dropping the highly reddened SNe 1989B and 1998bu in the
analysis, the change in the mean value of the absolute magnitudes
at maximum is minor (i.e., $\lesssim$0.03 mag). If the metallicity
effect was not taken into account, the distances to SN Ia host
galaxies and consequently the peak luminosity of SNe Ia would be
underestimated by $\sim$0.1 mag. Meanwhile, the scatter of the
mean absolute magnitudes would increase significantly, i.e. from
$\sim$0.11 mag to 0.16 mag in the $V$ band. This also argues in
favor of the metallicity correction of the Cepheid distances
purely by virtues of reducing the luminosity dispersion of the SN
calibrators

Inspection of the $M^{0}$ values listed in Table 3 shows that the
Hubble flow SNe Ia have approximately the same absolute magnitudes
as the more nearby ones calibrated via Cepheids. This fact
suggests that the value of $H_{0}$ (72 km s$^{-1}$ Mpc$^{-1}$),
assumed for determining the distances to the SNe in the Hubble
flow, must have been very close to the $H_{0}$ value implied from
the KP Cepheid distances. This similarity is not a foregone
consclusion but results from a coincidence. If we take the
distance moduli from STS to calibrate SNe Ia, however, the
predicted absolute magnitudes of SNe Ia would increase by
$\sim$0.25 mag, i.e., $M_{B} = -19.58\pm0.06$ and $M_{V} =
-19.51\pm0.05$ mag, which will lead to a Hubble constant lower
than the assumed value.

Combinations of the peak absolute magnitudes M$^{corr}_{UBVI}$
shown in Eqs. (12)-(15) with the individual apparent magnitudes
m$^{corr}_{UBVI}$ of 109 SNe Ia can give the luminosity distances
to 109 SNe Ia [Table 2, columns (6)-(9)]. The luminosity
distances, especially those in the $U$ band, are important for
comparison between the cosmological distances of SNe Ia beyond
z$\geq$0.8. The distance moduli presented in Table 2 can also be
used to constrain the peculiar motions of the nearby galaxies
hosting SNe Ia.

\subsection{Determination of the Hubble constant H$_{0}$}
Taking the zero-points of the Hubble lines from Table 4 and
inserting the absolute magnitudes of the SN calibrators into
Eq.(11), we can obtain the value for the Hubble constant $H_{0}$.
The key results for five different combinations of the Hubble flow
SNe Ia and the nearby calibrators are summarized in Table 6.

For all of the Hubble flow SNe (N = 73) and the nearby calibrators
(N = 11), the combination yields consistent determinations of
$H_{0}$ = 72$\pm$2 in the $B$, $V$, and $I$ bands; the smaller
value yielded in the $U$ band may be due to the statistical
fluctuation. Omitting 14 peculiar SNe from the Hubble flow and 2
from the nearby calibrators, the similar procedure reproduced well
the $H_{0}(U, B, V, I)$ values for the full sample. By further
eliminating the highly reddened SNe [i.e., those with E$(B -
V)_{host}$ $>$ 0.15 mag], one finds similar $H_{0}(U,B,V,I)$
values from the remaining sample (47+7). We also examined the
extreme case for the spectroscopically peculiar SNe Ia. For
91T/99aa-like SNe Ia, the resulting $H_{0}$ value is 68$\pm$3,
while the 91bg-like events give a value of 75$\pm$5. It is
reassuring that both the normal SNe Ia and the peculiar ones yield
consistent determinations of the Hubble constant within the
errors.

In all of the above five cases, the weighted mean value of $H_{0}$
is within 68$\leq H_{0}\leq$75. This suggests that our results are
not sensitive to different subsets of SNe Ia, and that no obvious
systematic errors are being introduced by absorption corrections
or including the peculiar SNe in the analysis. Taking the former
three cases shown in Table 6 as the most consistent
determinations, a value of 72$\pm$1 is obtained for the Hubble
constant. Restricting the analysis to the four best calibrators of
SNe 1981B, 1990N, 1994ae, and 1998aq as ranked by Riess et al
(2005), the resulting $H_{0}$ value remains almost unchanged.
Taking into account the intrinsic luminosity dispersion implied
from the Hubble flow SNe Ia, e.g., $\lesssim$0.12 mag in $BVI$
bands, the $H_{0}$ value becomes as
\begin{equation}
H_{0} = 72 \pm 4 (statistical) km s^{-1} Mpc^{-1}
\end{equation}
The uncertainty accounts for the statistical error of the absolute
magnitude calibration of nearby calibrators and the dispersion in
the Hubble diagram of Hubble flow SNe Ia.

In accord with Riess et al. (2005), especially their Table 13 (but
also see Freedman et al. 2001 for more detailed discussions of the
systematic error budget), two main sources of error were
incorporated into the systematic error budget. Uncertainties in
the LMC zero point enters at a level of 0.10 mag and the slope of
the $P-L$ relation is at a level of 0.05 mag. In quadrature, the
overall systematic error budget amounts to 0.11 mag, corresponding
to about 6\% in $H_{0}$. In combination with the statistical error
and the systematic error, the final result for the Hubble constant
yields
\begin{equation}
H_{0} = 72 \pm 4 (statistical) \pm 4 (systematic) = 72 \pm 6
(total) km s^{-1} Mpc^{-1}
\end{equation}
This $H_{0}$ value, based on the homogenization of the $\Delta
C_{12}$ method, is in excellent agreement with that determined by
the MLCS2k2 method (Riess et al. 2005), and is also fully
consistent with the determinations from the Wilkinson Microwave
Anisotropy Probe (WMAP) data (Spergel et al. 2003). However, the
true uncertainty in our value of the Hubble constant may indeed be
somewhat larger than our formal value due to the controversial
Cepheid distances to the SN Ia host galaxies (e.g., Saha et al.
2006).

\section{Discussion and Conclusions}
A larger sample of 109 SNe Ia has been compiled from the
literature to investigate the properties of the dust in their host
galaxies. Using the host galaxy reddening derived from $\Delta
C_{12}$ and the tail colors of SNe Ia, we found smaller values for
the reddening ratios of R$_{U}$ = 4.37$\pm$0.25, R$_{B}$ =
3.33$\pm$0.11, R$_{V}$ =2.30$\pm$0.11, and R$_{I}$ =
1.18$\pm$0.11, which are smaller than the standard R$_{UBVI}$
values of 5.5, 4.3, 3.3, and 1.8, respectively. The drastic
explosion of the SNe may in some manner change the distribution or
properties of the dust grains surrounding the pre-supernova.
Another possible explanation for the observed lower values of $R$
turns to the dust in the circumstellar environment of SNe (Wang
2005), which may be substantially different from the interstellar
dust.

In particular, we have discussed the luminosity dependence on the
environmental parameters, such as the morphological type of the
host galaxy and the location of the SNe within the galaxy. As
first noted by Hamuy et al. (2000) and also confirmed by our
study, the brighter SNe prefer to occurring in the late-type
(spiral) galaxies and the fainter ones prefer to occurring in the
early-type (E/S0) galaxies. However, there are two counter
examples: one is SN 1998es, which was a 91T-like event and
exploded in the lenticular galaxy NGC 632; the other is SN 1999by,
which was a 91bg-like event and occurred in the Sb$-$type galaxy
NGC 2841. These two cases suggest that the age may not be an
exclusive factor underlying SN Ia diversities. The radial gradient
of the absolute magnitudes (at a confidence of $\sim2.5\sigma$)
found for normal SNe Ia in E/S0 galaxies implies that the
metallicity is probably another important factor responsible for
the range of SN Ia luminosities.

Using 73 Hubble flow SNe Ia, we also examined the correlations
between SN Ia luminosities and the secondary parameters such as
the decline rate $\Delta m_{15}$, the peak $B - V$ color, and the
postmaximum color $\Delta C_{12}$. We found that the relation
between the peak luminosity and the light curve shapes becomes
apparently nonlinear, and more complicated function is needed to
describe this relationship. The correlation of the peak luminosity
and the peak color shows a large scatter of $\sim$0.23 mag around
the best fit line, which is not tight enough to get reliable
calibrations for SNe Ia. Compared with $\Delta m_{15}$ and
$B_{max} - V_{max}$, the color parameter $\Delta C_{12}$ can
depict extremely well the peak luminosity of SNe Ia, with an
impressively small dispersion of $\lesssim$0.12 mag in $BVI$. One
example to illustrate the superiority of $\Delta C_{12}$
calibration is SN 2001ay, which has a very broad light curve with
$\Delta m_{15}$ = 0.69 but a normal luminosity, i.e. M$_{V}$ =
$-$19.0 mag. If a $\Delta m_{15}$ correction is applied, the
magnitude would be overcorrected by $\sim$0.8 mag, i.e.
$\sim$$-$18.5 mag. Whereas the $\Delta C_{12}$ correction gives a
more reasonable magnitude of $-$19.10 mag, much closer to the
fiducial value: $M_{V}\approx-19.25$ for SN 1992al. Note that the
$\Delta C_{12}$ method, at present, is still an officially
empirical relation and more is needed to understand its underlying
physics, e.g., the well-behaved opacity in the expanding fireball.

The Hubble diagrams in $U$, $B$, $V$, and $I$ are displayed for 73
Hubble flow SNe Ia using the fully corrected apparent magnitudes.
Solutions for the zeropoint $\alpha$ of the Hubble line are given
for various subsets with different restrictions. We first
inspected the velocity restriction. The Hubble diagram constructed
by the two subsets, with different velocities demarcated at $v$ =
7,000 km s$^{-1}$ yet comparable sizes of SN sample, show little
variation in $\alpha$ values and the scatters. Inspections of the
reddening restriction suggests that the $\alpha$ values are also
insensitive to the accepted maximum value of a host galaxy
reddening. The uncertainties caused by the absorption correction
were found to be $\lesssim$0.08 mag in $BVI$, which justifies the
reliability of our adopted absorption correction for SNe Ia. We
finally examined the effect of the peculiar SNe Ia on the Hubble
line. After $\Delta C_{12}$ correction, the larger luminosity
difference between the normal SNe Ia and the peculiar ones has
been narrowed down to $\pm$0.15 mag. Inclusion of a fraction of
$\lesssim$20\% peculiar SNe Ia in the Hubble diagram will not
significantly change the Hubble line.

With Cepheid distances to 11 nearby SN Ia, including two peculiar
ones SNe 1991T and 1999by, we have calibrated the absolute
magnitudes of SNe Ia and found the fully corrected mean value:
$M^{0}_{U} = -19.89\pm0.08$, $M^{0}_{B} = -19.33\pm0.06$,
$M^{0}_{V} = -19.27\pm0.05$, and $M^{0}_{I} = -18.97\pm0.04$. The
forthcoming observations of SN 2006X in M100 (NGC 4321, which has
a Cepheid distance; see Freedman et al. 2001) will further improve
the determinations of the absolute magnitudes of SNe Ia. Applying
the above calibration value to the Hubble flow SNe Ia, we derived
the Hubble constant to be 72 $\pm$ 6 (total) km s$^{-1}$
Mpc$^{-1}$. This value seems to be more robust and does not change
with various combinations of distant SNe Ia and nearby
calibrators. Reducing the uncertainty in the Hubble constant
relies on a better understanding of the Cepheid $P-L$ relation,
the metallicity correction, and more importantly, a more precise
estimate of the distance to the LMC.

\acknowledgments This work has been supported by the National
Science Foundation of China (grants 10303002), the Basic Research
Foundation at Tsinghua University (JCqn2005036), and the National
Key Basic Research Science Foundation (NKBRSF TG199075402). We
thank the referee for a number of valuable comments and
suggestions that helped us to improve the paper. We are grateful
to Dr. Weidong Li for allowing us to use the data of SN 2002el
before publication.

\clearpage

\setlength\oddsidemargin   {-1.45cm}%
\setlength\evensidemargin  {-1.45cm}%
\setcounter{table}{0}
\def\baselinestretch{1.0}
\begin{deluxetable}{llcccccccccccc}
\tablewidth{580pt} \tabletypesize{\tiny} \tablecaption{Parameters
of well-observed nearby SNe\,Ia.\label{tab:SN1}}
\tablehead{
 \colhead{SN} &
 \colhead{Galaxy} &
 \colhead{$\!\!v_{\rm CMB/220}\!\!$}&
 \colhead{$U_{\max}$} &
 \colhead{$B_{\max}$} &
 \colhead{$V_{\max}$} &
 \colhead{$I_{\max}$} &
 \colhead{$\!\Delta m_{15}\!$} &
 \colhead{$\!\Delta C_{12}\!$}&
 \colhead{$E(B\mbox{-}V)$}&
 \colhead{$E(B\mbox{-}V)$} &
 \colhead{$E(B\mbox{-}V)$} &
 \colhead{$E(B\mbox{-}V)$} &
 \colhead{Ref.} \\
 \colhead{}  & \colhead{}  &
 \colhead{}  & \colhead{}  &
 \colhead{}  & \colhead{}  &
 \colhead{}  & \colhead{} &
 \colhead{}  & \colhead{Gal} &
 \colhead{tail} & \colhead{$\Delta C_{12}$}&
 \colhead{host}& \colhead{}\\
 \colhead{(1)}  & \colhead{(2)}  &
 \colhead{(3)}  & \colhead{(4)}  &
 \colhead{(5)}  & \colhead{(6)}  &
 \colhead{(7)}  & \colhead{(8)}  &
 \colhead{(9)}  & \colhead{(10)} &
 \colhead{(11)} & \colhead{(12)} &
 \colhead{(13)} & \colhead{(14)}
}
\startdata
\multicolumn{14}{c}{}\\
\noalign{\smallskip}
\tableline
1937C     &  IC  4182    &  330 & \nodata  &   8.74(09)   &  8.77(11)  &\nodata   &0.87(10) &  0.21(05)&  0.014 & 0.04(05)&0.03(06)&0.04(04)&1,2\\
1972E     &  NGC 5253    &  167 & 7.44(20) &    8.11(10)  &  8.17(09)  &\nodata   &0.87(10) &  0.24(05)&  0.056 & 0.05(05)&0.06(06)&0.05(04)&3\\
1974G     &  NGC 4414    &  655 & \nodata  &   12.45(16)  &  12.34(19) &\nodata   &1.03(10) &  0.47(07)&  0.018 & 0.07(17)&0.17(08)&0.15(07)&4\\
1981B     &  NGC 4536    &  1179& 11.72(12)&  11.96(08)   &  11.92(07) &\nodata   &1.11(07) &  0.47(05)&  0.018 & 0.10(14)&0.13(06)&0.12(05)&5\\
1986G     &  NGC 5128    &  317 & 12.57(12)&  11.95(08)   &  11.07(07) &\nodata   &1.73(05) &  1.49(05)&  0.115 & 0.54(06)&0.60(06)&0.57(04)&6,7\\
1989B     &  NGC 3627    &  549 & 12.15(11)&  12.20(07)   &  11.88(06) &11.64(06) &1.35(05) &  0.82(04)&  0.032 & 0.48(16)&0.41(06)&0.42(06)&8\\
1990N     &  NGC 4639    &  1179& 12.24(09)&  12.64(07)   &  12.64(05)&12.89(04)&1.06(03)&  0.36(02)&  0.026 & 0.16(05)&0.05(05)&0.11(04)&9\\
1990O     &  MCG 3-44-03 &  9175 & \nodata &16.19(10)&  16.22(08)  &  16.56(09)&  0.96(10)&  0.32(03)&  0.093&0.07(06)&0.04(05)&0.05(04)&10\\
1990af    &  Anon 213562 &  14966& \nodata &  17.77(07)  &  17.75(06) &\nodata  & 1.62(05)&  0.62(04)&  0.035&0.00(10)&$-$0.02(06)&0.00(04)&10\\
1991T$^{\dag}$&  NGC 4527&  1179 &11.16(09)&11.60(07)&  11.44(06)  &  11.63(05)&  0.95(05)&  0.48(05)&  0.022&0.15(05)&0.22(06)&0.18(04)&9\\
1991ag    &  IC 4919     &  4161 & \nodata &14.35(14)&  14.36(16)  &  14.68(16)&  0.88(10)&  0.22(05)&  0.062&0.09(07)&0.01(07)&0.05(05)&10\\
1991bg$^{\ddag}$&NGC 4374&  1179 & \nodata &14.58(08)&  13.83(07)  &  13.43(06)&  1.93(05)&  1.51(08)&  0.040&0.00(09)&0.06(08)&0.04(07)&10\\
1992A     & NGC 1992A    &  1338 & \nodata &12.50(07)&  12.50(07)  &  12.77(06)&  1.47(05)&  0.51(03)&  0.018&0.00(03)&0.02(04)&0.02(03)&10,11\\
1992P     & IC 3690      &   7939&\nodata&  16.05(07)&  16.09(06)  &  16.38(08)&  0.87(10)&  0.22(03)&  0.021 &0.10(05)&0.03(05)&0.07(04)&10\\
1992ae    &  Anon 2128-61&  22366& \nodata &18.57(10)&  18.43(08)  &  \nodata  &  1.47(10)&  0.51(05)&  0.036&0.04(11)&0.02(06)&0.02(05)&10\\
1992ag    &  ESO 508-G67 &  8095 & \nodata &16.23(08)&  16.16(07)  &  16.38(06)&  1.10(10)&  0.45(05)&  0.097&0.45(20)&0.12(06)&0.15(07)&10\\
1992al    &  ESO 234-G69 &  4214 &\nodata  &14.44(07)&  14.54(06)  &  14.88(06)&  1.11(05)&  0.31(03)&  0.034&0.05(05)&$-$0.02(05)&0.00(04)&10\\
1992aq    &  Anon 2304-37&  30014&\nodata  &19.31(12)&  19.29(08)  &  19.46(12)&  1.33(10)&  0.49(05)&  0.012&0.03(26)&0.09(07)&0.09(07)&10\\
1992bc    &  ESO 300-G9  &  5876 &\nodata  &15.04(07)&  15.17(06)  &  15.51(05)&  0.87(05)&  0.12(03)&  0.022&$-$0.03(05)&$-$0.07(05)&0.00(04)&10\\
1992bg    &  Anon 0741-62&  10936&\nodata  &16.59(08)&  16.67(07)  &  16.98(06)&  1.09(10)&  0.29(04)&  0.185&0.05(05)&$-$0.03(06)&0.01(04)&10\\
1992bh    &  Anon 0459-58&  13519&\nodata  &17.64(08)&  17.53(06)  &  17.71(07)&  1.07(10)&  0.51(05)&  0.022&0.12(08)&0.19(07)&0.18(05)&10\\
1992bk    &  ESO 156-G8  &  17237&\nodata&  18.01(12)&  18.05(12)  &  18.24(11)&  1.60(10)&  0.62(08)&  0.015&$-$0.01(10)&$-$0.01(09)&0.00(05)&10\\
1992bl    &  ESO 291-G11 &  12661&\nodata&  17.30(08)&  17.32(07)  &  17.57(06)&  1.49(10)&  0.51(05)&  0.011&0.04(06)&0.02(07)&0.03(04)&10\\
1992bo    &  ESO 352-G57 &  5445&\nodata&   15.74(07)&  15.76(06)  &  15.92(05)&  1.69(05)&  0.70(03)&  0.027&$-$0.01(12)&$-$0.07(05)&0.00(05)&10\\
1992bp    & Anon 0336-18 &  23557&\nodata&  18.29(07)&  18.30(06)  &  18.58(07)&  1.32(10)&  0.38(06)&  0.069&0.01(21)&$-$0.00(07)&0.00(07)&10\\
1992br    & Anon 0145-56 &  26259&\nodata&  19.31(17)&  19.19(10)  &  \nodata  &  1.66(10)&  0.71(06)&  0.026&0.03(23)&$-$0.01(08)&0.00(10)&10\\
1992bs    & Anon 0329-37 &  18787&\nodata&  18.33(09)&  18.25(07)  &  \nodata  &  1.34(10)&  0.50(04)&  0.011&0.07(10)&0.11(06)&0.09(05)&10\\
1993B     & Anon 1034-34 &  21011&\nodata&  18.37(11)&  18.39(09)  &  18.55(11)&  1.20(10)&  0.46(05)&  0.079 &0.27(18)&0.10(07)&0.12(07)&10\\
1993H     & ESO 445-G66  &  7523 &\nodata&  16.73(07)&  16.52(06)  &  16.51(06)&  1.69(10)&  0.82(05)&  0.060 &0.06(05)&0.04(07)&0.05(04)&10\\
1993O     & Anon 1331-33 &  15867&\nodata&  17.56(07)&  17.66(06)  &  17.89(06)&  1.26(05)&  0.41(03)&  0.053 &0.05(07)&0.03(05)&0.04(04)&10\\
1993ac    & PGC 17787    &  14674&\nodata&  17.75(16)&  17.71(12)  &  17.83(11)&  1.43(10)&  0.55(06)&  0.163 &$-$0.02(10)&0.09(07)&0.05(06)&12\\
1993ag    & Anon 1003-35 &  15013&\nodata&  17.81(08)&  17.73(06)  &  18.03(06)&  1.36(10)&  0.57(05)&  0.112 &0.12(08)&0.15(07)&0.14(05)&10\\
1994D     & NGC 4526     &  1179 &11.18(09)&11.75(07)&  11.83(06)  &  12.11(05)&  1.27(05)&  0.34(03)&  0.022 &0.05(08)&$-$0.03(05)&0.00(04)&13,14\\
1994M     & NGC 4493     &  7289 &\nodata&  16.24(08)&  16.22(07)  &  16.35(06)&  1.44(10)&  0.54(04)&  0.023 &0.13(06)&0.07(06)&0.10(04) &12\\
1994S     & NGC 4495     &  4829 &\nodata&  14.71(07)&  14.77(06)  &  15.08(06)&  1.02(10)&  0.26(05)&  0.021 &0.00(10)&0.04(07)&0.00(07)&12\\
1994T     & PGC 46640    &  10703&\nodata&  17.34(09)&  17.16(08)  &  17.35(07)&  1.55(10)&  0.69(10)&  0.029 &0.07(20)&0.13(07)&0.12(07)&12\\
1994ae    & NGC 3370     &  1575 &12.28(12)&12.89(07)&  12.94(06)  &  13.25(05)&  0.89(05)&  0.22(02)&  0.031 &0.03(05)&0.00(05)&0.02(04)&15\\
1995D     & NGC 2962     &  2129 &\nodata&  13.18(07)&  13.23(06)  &  13.58(05)&  1.00(05)&  0.31(03)&  0.058 &0.09(05)&0.02(05)&0.06(03)&12\\
1995E     &NGC 2441      &   3510 &\nodata&  16.70(07)&  16.01(06)  &  15.28(05)&  1.15(05)&  1.09(05)&  0.027 &0.73(15)&0.74(07)&0.74(06)&12\\
1995ac$^{\dag}$&Anon 2245-08&14635&\nodata& 17.04(07) &  17.07(06)  &  17.28(06)&  0.91(05)&  0.21(05)&  0.042 &0.04(07)&$-$0.03(06)&0.00(05)&12\\
1995al    &NGC 3021      &   1851 &\nodata&  13.26(07)&  13.22(06)  &  13.50(05)&  0.95(05)&  0.37(03)&  0.014 &0.18(05)&0.12(05)&0.15(04)&12\\
1995ak    &IC 1844       &   6589 &\nodata&  16.00(08)&  15.91(09)  &  16.12(09)&  1.35(10)&  0.51(03)&  0.038 &0.20(12)&0.10(05)&0.11(05)&12\\
1995bd$^{\dag}$&UGC 3151 &   4326 &\nodata&  15.16(07)&  14.92(06)  &  15.03(06)&  1.01(05)&  0.54(03)&  0.498 &0.18(05)&0.24(05)&0.21(04)&12\\
1996C     &MCG 8-25-47   &   9036&\nodata&  16.48(10)&  16.49(10)  &  16.65(08)&  0.94(10)&  0.34(05)&  0.013 &0.08(06)&0.09(07)&0.08(05)&12\\
1996X     &NGC 5061      &   2120&\nodata&  12.96(07)&  12.97(06)  &  13.23(05)&  1.33(05)&  0.44(02)&  0.069 &0.06(05)&0.04(04)&0.05(03)&12,16\\
1996Z     &NGC 2935      &   2285&\nodata&  14.32(07)&  14.00(06)  &  \nodata  &  1.06(10)&  0.68(05)&  0.064 &\nodata &0.37(06)&0.37(06)&12\\
1996ab    &Anon 1521+28  &   37370&\nodata&  19.54(09)&  19.43(08)  &  \nodata  &  1.16(05)&  0.38(10)&  0.032 &0.03(08)&0.05(10)&0.04(07)&12\\
1996ai    &NGC 5005      &   1298 &\nodata&  16.90(08)&  15.17(09)  &  13.93(10)&  1.02(10)&  1.92(05)&  0.014 &2.03(14)&1.63(06)&1.69(06)&12\\
1996bk    &NGC 5308      &   2462 &\nodata&  14.70(09)&  14.40(11)  &  14.26(08)&  1.69(10)&  1.00(07)&  0.018 &0.31(13)&0.23(08)&0.26(07)&12\\
1996bl    &Anon 0036+11  &   10447&\nodata&  16.68(07)&  16.68(07)  &  16.87(07)&  1.11(10)&  0.36(04)&  0.092 &0.13(07)&0.03(06)&0.07(04)&12\\
1996bo    &NGC 673       &   4898 &\nodata&  15.82(07)&  15.53(06)  &  15.52(06)&  1.30(05)&  0.72(03)&  0.077 &0.30(10)&0.34(05)&0.33(04)&11,12\\
1996bv    &UGC 3432      &   5016&\nodata&  15.34(13)&  15.14(10)  &  15.35(09)&  0.95(10)&  0.44(06)&  0.105&0.21(10)&0.18(07)&0.19(05)&12 \\
1997E     &NGC 2258      &   3998&14.74(09)&15.05(08)&  15.02(07)  &  15.17(06)&  1.44(05)&  0.56(05)&  0.124 &0.07(05)&0.10(07)&0.09(04)&17\\
1997Y     & NGC 4675     &   4968&14.82(10)& 15.26(08)&  15.24(08)  &  15.38(06)&  1.26(10)&  0.47(05)&  0.017 &0.08(05)&0.10(06)&0.09(04)&17\\
1997bp    &NGC 4680      &   2647&13.84(09)&13.89(07)&  13.72(06)  &  14.03(05)&  1.19(06)&  0.58(05)&  0.044 &0.14(05)&0.22(07)&0.17(04)&11,17\\
1997bq    & NGC 3147     &   2879&14.10(10)&14.44(11)&  14.24(08)  &  14.39(10)&  1.11(10)&  0.54(05)&  0.032 &0.17(05)&0.24(07)&0.19(04)&17\\
1997br$^{\dag}$&ESO 579-G40& 2193&13.00(15)&13.61(10)&  13.40(08)  &  13.42(20) &  1.15(10)& 0.75(03)&  0.113 &0.31(05)&0.41(05)&0.36(04)&17,18\\
1997cn$^{\ddag}$& NGC 5490&  5246&16.89(20)&16.89(15)&  16.35(10)  &  16.21(10)&  1.88(10)&  1.28(06)&  0.027 &0.06(13)&0.00(07)&0.01(06)&17,19\\
1997dg    &Anonymous     &   9845&16.31(10)&16.83(07)&  16.84(07)  &  16.92(06)&  1.19(10)&  0.44(05)&  0.078 & 0.29(20)&0.09(06)&0.11(07)&17\\
1997do    &UGC 3845      &   3135&14.00(15)&14.32(11)&  14.28(07)  &  14.52(07)&  0.99(10)&  0.40(06)&  0.063 &0.16(08)&0.13(08)&0.15(06)&17\\
1998V     & NGC 6627     &   5161&14.53(10)& 15.06(08)&  15.05(07)  &  15.27(06)&  1.08(10)&  0.35(04)&  0.196 &0.08(05)&0.03(06) &0.06(04)&17\\
1998ab$^{\dag}$&NGC 4704 &   8354&15.59(12)& 16.05(09)&  16.06(08)  &  16.29(08)&  1.12(10)&  0.39(03)&  0.017 &0.18(07)&0.05(05)&0.09(04)&17\\
1998aq    & NGC 3982     &   1514&11.63(09)& 12.30(07)&  12.41(06)  &  12.69(05)&  1.05(03)&  0.28(03)&  0.014 &$-$0.03(05)&0.07(04)& 0.03(04)&17\\
1998bp$^{\ddag}$&NGC 6495&   3048&15.21(10)& 15.32(10)&  15.02(08)  &  14.91(08)&  1.83(05)&  1.06(05)&  0.076 &0.06(05)&$-$0.03(07)&0.02(04)&17\\
1998bu    & NGC 3368     &   810&11.77(09) & 12.11(07)&  11.80(06)  &  11.59(05)&  1.02(03)&  0.64(03)&  0.025 &0.37(05)&0.34(05)&0.36(04)&17\\
1998de$^{\ddag}$& NGC 252&   4713&17.90(20)& 17.31(07)&  16.63(06)  &  16.48(08)&  1.95(09)&  1.58(06)&  0.058 &0.00(10)&0.03(07)&0.02(06)&17,20\\
1998dh    & NGC 7541     &   2766&13.53(20)& 13.81(08)&  13.74(07)  &  13.92(07)&  1.28(10)&  0.53(06)&  0.068 &0.15(05)&0.15(07)&0.15(04)&17\\
1998dx    & UGC 11149    &  14895&17.03(16)& 17.53(12)&  17.64(10)  &  17.75(09)&  1.47(10)&  0.43(05)&  0.041 &$-$0.04(08)&$-$0.05(07)&0.00(06)&17\\
1998eg    & UGC 12133    &   7067&15.63(16)& 16.07(07)&  16.07(06)  &  16.24(06)&  1.13(05)&  0.44(04)&  0.123 &0.08(10)&0.10(06)&0.09(06)&17\\
1998es$^{\dag}$ & NGC 632&   2872&13.28(09)& 13.81(07)&  13.73(06)  &  14.00(05)&  0.88(05)&  0.33(04)&  0.032 &0.17(06)&0.13(06)&0.15(05)&17\\
1999aa$^{\dag}$ & NGC 2595&   4572&14.15(09)& 14.71(07)&  14.73(06) &  15.13(05)&  0.82(05)&  0.19(03)&  0.040 &0.04(05)&0.05(05)&0.05(04)&17\\
1999ac$^{\dag}$ & NGC 6063&   2950&13.77(09)& 14.07(09)&  14.05(07) &  14.22(06)&  1.35(05)&  0.58(05)&  0.046 &0.07(07)&0.17(07)& 0.12(06)&17,21\\
1999aw$^{\dag}$ & Anon 1101-06&12363&\nodata& 16.72(07)&  16.70(06) &  17.17(05)&  0.81(03)&  0.16(03)&  0.032 &0.00(10)&0.04(05)&0.04(05)&17,22\\
1999by$^{\ddag}$& NGC 2841&   896&13.73(09)& 13.59(07)&  13.10(06)  &  12.88(05)&  1.90(05)&  1.29(06)&  0.016 &0.00(04)&$-$0.06(07)&0.00(03)&17,23\\
1999cc    & NGC 6038     &   9461&16.56(10)& 16.77(10)&  16.74(09)  &  16.88(09)&  1.45(10)&  0.54(04)&  0.023 &0.02(06)&0.07(06)&0.05(04)&17,24\\
1999cl    & NGC 4501     &   1179&15.57(10)& 14.86(12)&  13.73(08)  &  13.00(07)&  1.29(10)&  1.55(06)&  0.038 &1.24(06)&1.16(07)&1.20(05)&17,24\\
1999cp    & NGC 5468     &   3109&\nodata  & 13.94(09)&  13.96(08)  &  14.19(07)&  1.05(10)&  0.34(06)&  0.024 &\nodata&0.03(07)&0.03(07)&17\\
1999da$^{\ddag}$& NGC 6411&  3644&\nodata  & 16.61(09)&  16.03(09)  &  15.73(08)&  1.94(10)&  1.46(06)&  0.058 &0.05(20)&$-$0.04(08)&0.00(07)&24\\
1999dk    & UGC 1087     &   4184&14.47(13)& 14.81(09)&  14.76(07)  &  15.10(06)&  0.99(10)&  0.41(05)&  0.054 &0.15(21)&0.11(07)&0.11(06)&11,24\\
1999dq$^{\dag}$ & NGC 976&   4066&13.86(09)& 14.41(07)&  14.33(06)  &  14.57(05)&  0.95(05)&  0.34(05)&  0.110 &0.15(07)&0.07(07)&0.11(05)&11,17\\
1999ee    & IC 5179      &   3153&14.63(09)& 14.84(07)&  14.54(06)  &  14.62(05)&  0.94(05)&  0.54(03)&  0.020 &0.38(06)&0.29(05)&0.33(04)&24,25\\
1999ej    & NGC 495      &   3839&14.98(20)& 15.30(09)&  15.37(07)  &  15.53(06)&  1.39(10)&  0.47(05)&  0.072 &$-$0.02(17)&0.04(07)&0.02(06)&17\\
1999ek    & UGC 3329     &   5271&\nodata  & 15.53(07)&  15.42(05)  &  15.53(05)&  1.13(03)&  0.49(03)&  0.553 &0.20(08)&0.16(05)&0.17(04)&17,24\\
1999gd    & NGC 2623     &   5761&16.82(16)& 16.88(08)&  16.47(07)  &  16.19(08)&  1.13(10)&  0.82(06)&  0.041 &0.45(16)&0.48(07)&0.48(06)&17\\
1999gh    & NGC 2986     &   2314&13.88(20)& 14.21(16)&  14.13(12)  &  14.13(10)&  1.69(10)&  0.86(06)&  0.059 &0.07(05)&0.08(07)&0.07(04)&17\\
1999gp$^{\dag}$ &UGC 1993&   7811&15.47(10)& 16.01(08)&  15.95(07)  &  16.23(06)&  0.92(10)&  0.33(05)&  0.056 &0.11(10)&0.09(07)&0.10(06)&17\\
2000E     & NGC 6951     &   1824&12.50(10)& 12.72(07)&  12.63(06)  &  12.77(07)&  0.96(05)&  0.35(04)&  0.366 &0.25(10)&0.09(06)&0.13(05)&26\\
2000bh    & ESO 573-G014 &   7238&\nodata  & 15.86(08)&  15.89(07)  &  16.24(05)&  1.16(10)&  0.42(05)&  0.048 &0.07(06)&0.08(07)&0.07(05)&17\\
2000ca    & ESO 383-G032 &   7080&14.96(10)& 15.53(07)&  15.62(06)  &  15.93(05)&  0.96(05)&  0.22(03)&  0.067 &0.02(05)&$-$0.03(05)&0.00(04)&24\\
2000ce    & UGC 4195     &   4940&16.76(20) & 16.99(16)& 16.41(16)  &  16.04(12)&  1.00(10)&  0.85(06)&  0.057 &0.73(15)&0.56(07)&0.59(06)&17,24\\
2000cf    & MCG 11-19-25 &  10805&16.59(12)& 17.00(09)&  17.02(08)  &  17.21(07)&  1.27(10)&  0.44(04)&  0.032 &0.11(05)&0.06(05)&0.09(04)&17,24\\
2000cn    & UGC 11064    &   6971&16.47(10)& 16.56(08)&  16.40(07)  &  16.57(05)&  1.59(10)&  0.73(04)&  0.057 &0.12(06)&0.12(06)&0.12(05)&17\\
2000cx$^{\dag}$ & NGC 524&   2454&12.74(09)& 13.06(07)&  12.98(06)  &  13.49(06)&  0.93(05)&  0.32(05)&  0.083 &$-$0.21(05)&0.06(06)&0.00(04)&17,27\\
2000dk    & NGC 382      &   4929&15.04(09)& 15.33(07)&  15.34(06)  &  15.59(05)&  1.63(05)&  0.67(07)&  0.070 &$-$0.03(06)&0.00(08)&0.00(05)&17\\
2000fa    & UGC 3770     &   6526&15.39(20)& 15.70(10)&  15.72(10)  &  15.94(08)&  0.98(10)&  0.35(05)&  0.069 &0.12(08)&0.08(07)&0.10(06)&17\\
2001V$^{\dag}$& NGC3987  &   4804&\nodata  & 14.55(12)&  14.53(09)  &  14.82(08)&  0.95(05)&  0.28(04)&  0.020 &0.02(10)&0.01(06)&0.01(06)&29\\
2001ay$^{\ast}$ & IC 4423&   9269&\nodata  & 16.61(07)&  16.62(06)  &  16.67(06)&  0.69(05)&  0.38(04)&  0.019 &\nodata &\nodata &0.04(05)&28\\
2001ba    & MCG-05-28-001&   9245&\nodata  & 16.17(08)&  16.27(09)  &  16.58(07)&  0.98(05)&  0.27(05)&  0.064 &$-$0.01(10)&$-$0.01(07)&0.00(05)&24\\
2001bt    & IC 4830      &   4332&\nodata  & 15.25(07)&  15.10(05)  &  15.18(05)&  1.28(05)&  0.63(03)&  0.065 &0.26(06)&0.25(05)&0.25(04)&24\\
2001cn    & IC 4758      &   4628&14.99(18)& 15.20(07)&  15.11(05)  &  15.21(05)&  1.15(05)&  0.51(03)&  0.059 &0.21(07)& 0.16(05)&0.17(04)&24\\
2001cz    & NGC 4679     &   4900&\nodata  & 15.03(07)&  14.97(05)  &  15.14(05)&  1.07(03)&  0.36(03)&  0.092 &0.17(06)&0.04(05)&0.09(04)&24\\
2001el    & NGC 1448     &   1030&12.59(09)& 12.75(07)&  12.68(06)  &  12.79(06)&  1.13(03)&  0.58(02)&  0.014 &0.29(05)&0.24(05)&0.27(04)&24\\
2002bo    & NGC 3190     &   1547&\nodata  & 13.93(10)&  13.50(10)  &  13.47(10)&  1.13(05)&  0.74(05)&  0.025 &0.43(06)&0.40(07)&0.42(05)&30\\
2002cx$^{\ast}$& CGCG 044-035&7488&\nodata & 17.56(10)&  17.46(16)  &  17.29(10)&  1.28(10)&  0.60(05)&  0.032 &0.14(21)&0.20(07)&0.19(07)&27\\
2002el    & NGC 6986     &   7079&\nodata  & 16.11(07)&  16.16(06)  &  16.37(07)&  1.32(05)&  0.49(03)&  0.085 &0.28(18)&0.09(05)&0.10(05)&31\\
2002er    & UGC 10743    &   2652&13.86(09)& 14.21(07)&  14.07(06)  &  14.18(07)&  1.33(04)&  0.59(05)&  0.157 &0.21(06)&0.19(06)&0.20(06)&32\\
2002hu    & MCG+06-06-012&  11000&16.42(09)& 16.63(07)&  16.68(06)  &  16.93(07)&  1.03(05)&  0.31(04)&  0.044 &0.09(05)&0.01(05)&0.05(04)&33\\
2003du    & UGC 9391     &   2008&13.00(09)& 13.49(07)&  13.61(06)  &  13.83(07)&  1.00(04)&  0.22(05)&  0.010 &0.05(05)&$-$0.07(06)&0.00(04)&34\\
2004S     & MCG-05-16-021&   2516&\nodata  & 14.01(08)&  14.02(09)  &  14.32(10)&  1.11(05)&  0.40(05)&  0.101 &0.08(05)&0.07(06)&0.08(04)&35\\
\enddata

\tablerefs{[1]Schaefer(1994), [2]Pierce \& Jacoby (1995),
[3]Ardeberg \& Groot (1973), [4]Schaefer(1998), [5]Schaefer(1995),
[6]Phillips et al.(1987), [7]Cristiani et al.(1992), [8]Wells et
al. (1994), [9]Lira et al. (1998), [10]Hamuy et al. (1996a),
[11]Altavilla et al.(2004), [12]Riess et al. (1999), [13]Richmond
et al. (1995), [14]Patat et al. (1996), [15]Riess et al. (2005),
[16]Salvo et al. (2001), [17]Jha et al.(1999; 2006a), [18]Li et
al. (1999), [19]Turrato et al. (1998), [20]Modjaz et al. (2001),
[21]Phillips et al. (2006), [22]Strolger et al. (2002),
[23]Garnavich et al. (2004), [24]Krisciunas et~al. (2001, 2003,
2004a,b, 2005), [25]Stritzinger et al. (2002), [26]Valenini et al.
(2003), [27]Li et~al.(2001b, 2003), [28]Howell \& Nugent (2003),
[29]Vinko et al. (2003), [30]Benetti et al. (2004), [31]from
Weidong Li, [32]Pignata et~al.(2004), [33]Sahu et al. (2006), [34]
Anupama et al. (2005), [35]Misra et al. (2005).}

\tablenotetext{\dag}{1991T/1999aa-like SNe.}
\tablenotetext{\ddag}{1991bg-like SNe.} \tablenotetext{\ast}{SN
2001ay is the SN Ia with the broadest light curve (Howell \&
Nugent 2003); SN 2002cx displayed the most unique spectral and
photometric features (Li et al 2003).}
\end{deluxetable}

\clearpage
\clearpage
 \setlength\oddsidemargin   {-0.20cm}%
 \setlength\evensidemargin  {-0.20cm}%
\setcounter{table}{1}
\def\baselinestretch{1.1}
\begin{deluxetable}{llcrrccccc}
\tablewidth{0pt} \tabletypesize{\scriptsize}
\tablecaption{Parameters relevant for SN host
galaxies\label{tab:SN2}}
\tablehead{
 \colhead{SN} &
 \colhead{Galaxy} &
 \colhead{$z_{CMB/220}$} &
 \colhead{Type} &
 \colhead{T} &
 \colhead{r$_{SN}$/r$_{25}$} &
 \colhead{$\mu_{U}$} &
 \colhead{$\mu_{B}$} &
 \colhead{$\mu_{V}$} &
 \colhead{$\mu_{I}$} \\
%
%
 \colhead{(1)}  & \colhead{(2)}  &
 \colhead{(3)}  & \colhead{(4)}  &
 \colhead{(5)}  & \colhead{(6)}  &
 \colhead{(7)}  & \colhead{(8)}  &
 \colhead{(9)}  & \colhead{(10)}
 }
\startdata
\multicolumn{10}{c}{}\\
\noalign{\smallskip} \hline
1937C   &   IC  4182$^{\star}$  &   0.0011  &   Sm  &   9   &   0.28(03)    &   \nodata &   \nodata     &   \nodata     &   \nodata \\
1972E   &   NGC 5253$^{\star}$  &   0.0006  &   Im  &   8   &   \nodata &   \nodata    &   \nodata    &   \nodata    &   \nodata \\
1974G   &   NGC 4414$^{\star}$  &   0.0022  &   Sc  &   5   &   0.55(03)    &   \nodata &   \nodata    &   \nodata    &   \nodata \\
1981B   &   NGC 4536$^{\star}$  &   0.0039  &   Sbc &   4   &   0.70(05)    &   \nodata    &   \nodata    &   \nodata    &   \nodata \\
1986G   &   NGC 5128            &   0.0011  &   S0  &   $-$2    &   0.23(02)    &   28.36(24)   &   28.20(19)   &   28.15(17)   &   \nodata \\
1989B   &   NGC 3627$^{\star}$  &   0.0018  &   Sb  &   3   &   0.21(02)    &   \nodata    &   \nodata    &  \nodata    &   \nodata    \\
1990N   &   NGC 4639$^{\star}$  &   0.0499  &   Sbc &   4   &   0.84(06)    &  \nodata    &   \nodata   &   \nodata    &   \nodata   \\
1990O   &   MCG+3-44-03         &   0.0306  &   Sa  &   1   &   0.82(05)    &   \nodata &   35.43(20)   &   35.43(18)   &   35.51(16)   \\
1990af  &   Anon 213562         &   0.0039  &   Sa  &   1   &   0.00(05)    &\nodata &   36.50(19)   &   36.57(17)   &   9.99(99 \\
1991T   &   NGC 4527$^{\star}$  &   0.0039  &   Sbc &   4   &   0.61(02)    &  \nodata      &   \nodata     &   \nodata     &   \nodata \\
1991ag  &   IC  4919            &   0.0039  &   Sdm &   8   &   0.80(06)    &   \nodata &   33.78(25)   &   33.72(23)   &   33.73(21    \\
1991bg  &   IC  4374            &   0.0139  &   E1  & $-$4  &   0.30(02)    &   30.29(23)   &   30.35(19)   &   30.31(17)   &   30.40(14)   \\
1992A   &   NGC 1380            &   0.0045  &   S0  &   $-$2    &   0.44(03)    &   \nodata &   31.41(17)   &   31.46(16)   &   31.54(14)   \\
1992P   &   IC  3690            &   0.0265  &   Sb  &   3   &   0.89(13)    &   \nodata &   35.46(19)   &   35.43(17)   &   35.43(15)   \\
1992ae  &   Anon 2128-61        &   0.0746  &   E/S0    &   $-$3    &   0.15(05)    &   \nodata &   37.48(23)   &   37.39(19)   &   9.99(99 \\
1992ag  &   ESO 508-G67         &   0.0270  &   Sbc &   4   &   \nodata &   \nodata &   35.08(27)   &   35.10(22)   &   35.18(16)   \\
1992al  &   ESO 234-G69         &   0.0141  &   Sc  &   5   &   0.42(03)    &   \nodata &   33.77(19)   &   33.81(17)   &   33.85(14)   \\
1992aq  &   Anon 2304-37        &   0.1001  &   Sa  &   1   &   0.38(10)    &   \nodata &   38.17(29)   &   38.22(22)   &   38.23(19)   \\
1992bc  &   ESO 300-G9          &   0.0196  &   Sab &   4   &   0.80(23)    &   \nodata &   34.74(19)   &   34.71(17)   &   34.67(14)   \\
1992bg  &   Anon 0741-62        &   0.0364  &   Sa  &   1   &   0.18(10)    &   \nodata &   35.94(19)   &   35.96(17)   &   35.97(14)   \\
1992bh  &   Anon 0459-58        &   0.0451  &   Sbc &   4   &   0.11(02)    &   \nodata &   36.33(22)   &   36.36(18)   &   36.45(15)   \\
1992bk  &   ESO 156-G8          &   0.0575  &   E/S0    &   $-$2    &   0.87(08)    &   \nodata &   36.74(24)   &   36.87(21)   &   36.90(17)   \\
1992bl  &   ESO 291-G11         &   0.0422  &   S0/a    &   1   &   1.52(07)    &   \nodata &   36.20(19)   &   36.28(17)   &   36.33(14)   \\
1992bo  &   ESO 352-G57         &   0.0181  &   S0  &   $-$2    &   1.74(19)    &   \nodata &   34.31(22)   &   34.47(18)   &   34.50(15)   \\
1992bp  &   Anon 0336-18        &   0.0785  &   S0  &   $-$2    &   0.26(04)    &   \nodata &   37.48(27)   &   37.47(21)   &   37.48(16)   \\
1992br  &   Anon 0145-56        &   0.0875  &   E/S0    &   $-$3    &   1.23(07)    &   \nodata &   37.86(39)   &   37.88(28)   &   \nodata \\
1992bs  &   Anon 0329-37        &   0.0626  &   Sb  &   3   &   0.94(06)    &   \nodata &   37.17(22)   &   37.17(18)   &   \nodata \\
1993B   &   Anon 1034-34        &   0.0700  &   Sb  &   3   &   0.29(03)    &   \nodata &   37.24(28)   &   37.34(22)   &   37.35(18)   \\
1993H   &   ESO 445-G66         &   0.0251  &   Sab &   2   &   0.35(05)    &   \nodata &   35.00(19)   &   35.01(17)   &   34.96(14)   \\
1993O   &   Anon 1331-33        &   0.0529  &   S0  &   $-$2    &   1.40(40)    &   \nodata &   36.64(19)   &   36.75(17)   &   36.75(14)   \\
1993ac  &   PGC 17787           &   0.0489  &   E   &   $-$3    &   0.68(10)    &   \nodata &   36.55(28)   &   36.59(22)   &   36.55(18)   \\
1993ag  &   Anon 1003-35        &   0.0500  &   S0  &   $-$2    &   0.70(18)    &   \nodata &   36.44(22)   &   36.51(18)   &   36.71(15)   \\
1994D   &   NGC 4526            &   0.0243  &   S0  & $-$2 &0.19(01) &    30.99(23)   &   31.02(19)   &   31.06(17) & 31.05(14)\\
1994M   &   NGC 4493            &   0.0161  &   E   &   $-$4 &0.68(10) &    \nodata     &   34.99(19)   &   35.07(17)   &   35.07(14)   \\
1994S   &   NGC 4495            &   0.0357  &   Sab &   2   &0.74(09)  &   \nodata &   34.14(27)   &   34.11(21    &   34.10(16)   \\
1994T   &   PGC 46640           &   0.0052  &   Sa  &   1   &\nodata   &   \nodata &   35.77(28)   &   35.78(22)   &   35.92(16)   \\
1994ae  &   NGC 3370$^{\star}$  &   0.0039  &   Sc  &   5   &  0.51(04)    & \nodata &\nodata & \nodata & \nodata  \\
1995D   &   NGC 2962            &   0.0144  &   S0  &$-$1& 1.18(06)    &\nodata &   32.43(17)   &   32.45(16)   &   32.54(14) \\
1995E   &   NGC 2441            &   0.0117  &   Sb  &   3   &   0.41(03)    &   \nodata &   33.49(24)   &   33.52(20)   &   33.34(15)   \\
1995ac  &   Anon 2245-08        &   0.0071  &   Sa  &   1 & \nodata    &   \nodata &   36.56(22)   &   36.48(18)   &   36.35(15)   \\
1995ak  &   IC  1844            &   0.0488  &   Sbc & 4&   0.37(03)    &   \nodata &   34.79(22)   &   34.80(19)   &   34.87(16)   \\
1995al  &   NGC 3021            &   0.0062  &   Sbc &   4   &   0.44(05)    &   \nodata &   32.26(19)   &   32.27(17)   &   32.38(14)   \\
1995bd  &   UGC 3151            &   0.0220  &   Sbc &   4   &0.79(09)  & \nodata   &33.75(19)   & 33.68(17)  & 33.73(14) \\
1996C   &   MCG+8-25-47         &   0.0301  &   Sa  &   1   &   0.42(03)    &   \nodata &   35.64(23)   &   35.65(20)   &   35.58(16)   \\
1996X   &   NGC 5061            &   0.0071  &   E0  &   $-$5&   0.57(03)    &   \nodata &   31.97(17)   &   32.01(16)   &   32.06(14)   \\
1996Z   &   NGC 2935            &   0.0076  &   Sb  &   3   &   0.61(03)    &   \nodata &   32.42(24)   &   32.42(20)   &   \nodata \\
1996ai  &   NGC 5005            &   0.1246  &   Sbc &   4   &0.16(01)    &\nodata &   30.76(25)   &   30.67(21)   &   30.99(17)   \\
1996ab  &   Anon 1521+28        &   0.0043  &   Sc  &   5   &\nodata & \nodata &   38.68(28)   &   38.56(22)   &   \nodata   \\
1996bk  &   NGC 5308            &   0.0082  &   S0  &   $-$2&   \nodata &   \nodata &   32.33(28)   &   32.45(23)   &   32.49(17)   \\
1996bl  &   Anon 0036+11        &   0.0348  &   Sc  &   5   &   \nodata &   \nodata &   35.82(19)   &   35.82(17)   &   35.78(15)   \\
1996bo  &   NGC 673             &   0.0163  &   Sc  &   5   &   0.13(01)    &   \nodata &   33.90(19)   &   33.93(17)   &   34.02(14)   \\
1996bv  &   UGC 3432            &   0.0167  &   Scd &   6   &   0.07(01)    &   \nodata &   34.15(24)   &   34.06(20)   &   34.16(16)   \\
1997E   &   NGC 2258            &   0.0133  &   S0  &   $-$2    &   0.95(27)    &   33.81(23)   &   33.77(19)   &   33.85(17)   &   33.87(14)   \\
1997Y   &   NGC 4675            &   0.0166  &   Sb  &   3   &   0.14(02)    &   34.13(24)   &   34.15(19)   &   34.20(18)   &   34.17(14)   \\
1997bp  &   NGC 4680            &   0.0088  &   Sb  &   3   &   0.57(08)    &   32.72(23)   &   32.46(19)   &   32.46(17)   &   32.70(14)   \\
1997bq  &   NGC 3147            &   0.0096  &   Sbc &   4   &   0.62(03)    &   33.05(24)   &   33.06(21)   &   33.02(18)   &   33.10(16)   \\
1997br  &   ESO 576-G40$^{\ast}$ &   0.0073  &   Sd  &   7   &   0.90(06)    &   31.11(26)   &   31.59(20)   &   31.73(18)   &   31.89(24)   \\
1997cn  &   NGC 5490            &   0.0175  &   E   &   $-$5&   0.22(02)    &   34.19(35)   &   34.33(28)   &   34.21(21)   &   34.21(17)   \\
1997dg  &   Anonymous           &   0.0328  &   Sa  &   1   &   \nodata &   35.67(34)   &   35.75(27)   &   35.83(22)   &   35.74(16)   \\
1997do  &   UGC 3845            &   0.0104  &   Sbc &   4   &   0.12(01)    &   33.39(33)   &   33.27(26)   &   33.29(20)   &   33.37(16)   \\
1998V   &   NGC 6627            &   0.0172  &   Sb  &   3   &   0.86(10)    &   34.21(24)   &   34.23(19)   &   34.21(17)   &   34.19(14)   \\
1998ab  &   NGC 4704            &   0.0279  &   Sbc &   4   &   0.51(04)    &   35.11(24)   &   35.10(20)   &   35.14(18)   &   35.16(15)   \\
1998aq  &   NGC 3982$^{\star}$  &   0.0050  &   Sb  &   3   &   0.32(02)    &   \nodata    &   \nodata    &  \nodata    &   \nodata    \\
1998bp  &   NGC 6495            &   0.0102  &   E   &   $-$5    &   0.21(04)    &   33.08(24)   &   33.17(20)   &   33.19(18)   &   33.13(15)   \\
1998bu  &   NGC 3368$^{\star}$  &   0.0027  &   Sab &   2   &   0.24(01)    &   \nodata   &   \nodata    &   \nodata    &   \nodata    \\
1998de  &   NGC 252             &   0.0157  &   S0  &   $-$1    &   1.60(19)    &   34.39(35)   &   34.15(24)   &   34.05(20)   &   34.18(16)   \\
1998dh  &   NGC 7541            &   0.0092  &   Sbc &   4   &   0.52(02)    &   32.58(29)   &   32.50(19)   &   32.56(17)   &   32.64(15)   \\
1998dx  &   UGC 11149           &   0.0496  &   Sb  &   3   &   0.95(05)    &   36.60(33)   &   36.63(26)   &   36.74(21)   &   36.60(17)   \\
1998eg  &   UGC 12133           &   0.0236  &   Sc  &   6   &   0.70(07)    &   35.02(33)   &   35.02(24)   &   35.08(20)   &   35.06(15)   \\
1998es  &   NGC 632             &   0.0096  &   S0  &   $-$2&   0.23(03)    &   32.86(26)   &   32.89(22)   &   32.84(18)   &   32.92(15)   \\
1999aa  &   NGC 2595            &   0.0152  &   Sc  &   5   &   0.38(03)    &   34.27(23)   &   34.20(19)   &   34.13(17)   &   34.21(14)   \\
1999ac  &   NGC 6063            &   0.0098  &   Scd &   6   &   0.85(10)    &   32.74(30)   &   32.71(25)   &   32.83(20)   &   32.90(15)   \\
1999aw  &   Anon 1101-06        &   0.0412  &   ?   &   ?   &   \nodata &   \nodata &   36.29(22)   &   36.15(18)   &   36.28(15)   \\
1999by  &   NGC 2841$^{\star}$  &   0.0030  &   Sb  &   3   &   0.66(02)    &   \nodata    &   \nodata    &   \nodata    &   \nodata   \\
1999cc  &   NGC 6038            &   0.0315  &   Sc  &   5   &   0.50(07)    &   35.75(24)   &   35.58(20)   &   35.64(18)   &   35.61(16)   \\
1999cl  &   NGC 4501$^{\ast}$            &   0.0039  &   Sb  &   3   &   0.30(01)    &   30.12(27)   &   30.12(24)   &   30.18(19)   &   30.51(15)   \\
1999cp  &   NGC 5468            &   0.0104  &   Scd &   6   &   0.72(03)    &   \nodata &   33.17(28)   &   33.16(22)   &   33.12(16)   \\
1999da  &   NGC 6411            &   0.0121  &   E   &   $-$5&   1.03(030    &   \nodata &   33.71(28)   &   33.64(22)   &   33.55(17)   \\
1999dk  &   UGC 1087            &   0.0139  &   Sc  &   5   &   0.64(08)    &   33.91(32)   &   33.79(25)   &   33.79(20)   &   33.95(15)   \\
1999dq  &   NGC 976             &   0.0136  &   Sc  &   5   &   0.19(01)    &   33.48(27)   &   33.53(22)   &   33.46(18)   &   33.49(15)   \\
1999ee  &   IC 5179             &   0.0105  &   Sbc &   4   &   0.31(02)    &   33.34(23)   &   33.26(19)   &   33.19(17)   &   33.30(14)   \\
1999ej  &   NGC 495             &   0.0128  &   S0/a&   0   &   0.80(10)    &   34.41(35)   &   34.29(25)   &   34.39(20)   &   34.34(15)   \\
1999ek  &   UGC 3329            &   0.0176  &   Sbc &   4   &   0.70(12)    &   \nodata &   34.27(19)   &   34.28(16)   &   34.29(14)   \\
1999gd  &   NGC 2623            &   0.0192  &   Sab &   2   &   0.28(04)    &   34.54(33)   &   34.55(25)   &   34.59(20)   &   34.56(16)   \\
1999gh  &   NGC 2986            &   0.0077  &   E   &   $-$5&   0.57(04)    &   32.19(29)   &   32.38(24)   &   32.55(20)   &   32.54(16)   \\
1999gp  &   UGC 1993            &   0.0260  &   Sb  &   3   &   0.28(03)    &   35.14(31)   &   35.16(25)   &   35.11(20)   &   35.16(15)   \\
2000E   &   NGC 6951            &   0.0061  &   Sbc &   4   &   0.25(01)    &   32.06(27)   &   31.79(22)   &   31.73(18)   &   31.68(15)   \\
2000bh  &   ESO 573-G014        &   0.0241  &   Sc  &   5   &   0.50(22)    &   \nodata &   34.88(22)   &   34.94(18)   &   35.09(15)   \\
2000ca  &   ESO 383-G032        &   0.0236  &   Sbc &   4   &   0.25(04)    &   35.09(24)   &   35.03(19)   &   35.02(17)   &   34.99(14)   \\
2000ce  &   UGC 4195            &   0.0165  &   Sb  &   3   &   0.51(07)    &   34.21(35)   &   34.45(28)   &   34.40(25)   &   34.36(18)   \\
2000cf  &   MCG+11-19-25        &   0.0360  &   Sbc &   4   &   0.55(25)    &   35.98(24)   &   35.95(20)   &   36.03(18)   &   36.03(15)   \\
2000cn  &   UGC 11064           &   0.0232  &   Scd &   6   &   0.17(03)    &   35.04(27)   &   34.91(22)   &   34.96(18)   &   35.10(15)   \\
2000cx  &   NGC 524             &   0.0082  &   S0  &   $-$1&   \nodata &   32.60(23)   &   32.37(19)   &   32.24(17)   &   32.45(14)   \\
2000dk  &   NGC 382             &   0.0164  &   E   &   $-$5&   0.51(19)    &   33.98(26)   &   33.96(22)   &   34.09(18)   &   34.20(15)   \\
2000fa  &   UGC 3770            &   0.0217  &   Im  &   10  &   0.40(09)    &   35.00(35)   &   34.81(25)   &   34.85(21)   &   34.85(16)   \\
2001V   &   NGC 3987            &   0.0160  &   Sb  &   3   &   0.96(09)    &   \nodata &   33.92(26)   &   33.83(21)   &   33.82(16)   \\
2001ay  &   IC  4423            &   0.0309  &   Sbc &   4   &   0.69(10)    &   \nodata &   35.75(22)   &   35.75(18)   &   35.56(15)   \\
2001ba  &   MCG-05-28-001       &   0.0308  &   Sbc &   4   &   0.85(08)    &   \nodata &   35.58(22)   &   35.60(19)   &   35.59(15)   \\
2001bt  &   IC  4830            &   0.0144  &   Sbc &   4   &   0.57(05)    &   \nodata &   33.61(19)   &   33.69(16)   &   33.78(14)   \\
2001cn  &   IC  4758            &   0.0154  &   Sc  &   5   &   0.73(08)    &   34.06(28)   &   33.91(19)   &   33.95(16)   &   33.95(14)   \\
2001cz  &   NGC 4679            &   0.0163  &   Sc  &   4   &   0.44(04)    &   \nodata &   34.14(19)   &   34.09(16)   &   34.04(14)   \\
2001el  &   NGC 1448            &   0.0034  &   Scd &   6   &   \nodata &   31.30(23)   &   31.18(19)   &   31.33(17)   &   31.44(14)   \\
2002bo  &   NGC 3190            &   0.0052  &   Sa  &   1   &   0.17(01)    &   \nodata &   31.84(23)   &   31.79(20)   &   31.93(17)   \\
2002cx  &   CGCG 044-035        &   0.0250  &   ?   &   ?   &   \nodata &   \nodata &   36.06(28)   &   36.15(26)   &   35.94(18)   \\
2002el  &   NGC 6986            &   0.0236  &   E/S0&   $-$3&   0.86(07)    &   \nodata &   34.95(22)   &   35.08(18)   &   35.14(15)   \\
2002er  &   UGC 10743           &   0.0088  &   Sa  &   1   &   0.38(04)    &   32.67(30)   &   32.72(24)   &   32.76(20)   &   32.83(16)   \\
2002hu  &   MCG+06-06-012       &   0.0367  &   Sc  &   5   &   1.17(10)    &   36.22(23)   &   35.89(19)   &   35.91(17)   &   35.89(15)   \\
2003du  &   UGC 9391            &   0.0067  &   Sdm &   8   &   0.34(03)    &   33.13(23)   &   32.99(19)   &   33.01(17)   &   32.89(15)   \\
2004S   &   MCG-05-16-21        &   0.0084  &   Sc  &   5   &   1.13(05)    &   \nodata &   33.05(19)   &   33.09(18)   &   33.19(16)   \\
\enddata
\tablenotetext{\star}{Host galaxies for which a Cepheid distance
is available.} \tablenotetext{\ast}{The dust in ESO 576-G40 and
NGC 4501 may be very different from that of the other distant
galaxies (see the discussions in the text), so the distance moduli
presented here for them may not represent their true distances.}
\end{deluxetable}

\def\baselinestretch{1.1}
\begin{deluxetable}{lccccc}
\tablewidth{0pt} \tabletypesize{\footnotesize} \label{tab3}
\tablecaption{The coefficients of the relations between peak
luminosity, color parameter $\Delta C_{12}$ and the host-galaxy
reddening.}

\tablehead{
\colhead{$Bandpass$} &\colhead{N$^{a}$} & \colhead{$b$} &
\colhead{$R$} &\colhead{M$^{0}$}&\colhead{$\sigma_{M}$}}
\startdata
U&29&2.65(10)&4.37(25)&$-$19.87(05)&0.162\\
B&73&1.94(06)&3.33(11)&$-$19.34(02)&0.120\\
V&73&1.44(06)&2.30(11)&$-$19.27(02)&0.117\\
I&69&1.00(05)&1.18(11)&$-$18.96(02)&0.112\\
\enddata
\def\baselinestretch{1.5}
\tablenotetext{a}{N is the number of Hubble flow SN Ia sample used
to determine the relation.}
\end{deluxetable}

\def\baselinestretch{1.1}
\begin{deluxetable}{lccccccccc}
\tablewidth{0pt} \tabletypesize{\scriptsize} \label{tab4}
\tablecaption{The best-fitting results of the $m-z$ relation for
different SN Ia sample.} \tablehead{
 \colhead{Sample} & \colhead{$N^{a}_{BV}$} &
 \colhead{$\alpha_{U}$}&\colhead{$\sigma_{U}$}&\colhead{$\alpha_{B}$}&\colhead{$\sigma_{B}$}&
 \colhead{$\alpha_{V}$}&\colhead{$\sigma_{V}$}&\colhead{$\alpha_{I}$}&\colhead{$\sigma_{I}$}
}
\startdata
All &73&$-$4.18(03)&0.162&$-$3.64(01)&0.120&$-$3.56(01)&0.117&$-$3.25(01)&0.112\\
normal&58&$-$4.14(03)&0.113&$-$3.63(01)&0.104&$-$3.54(01)&0.099&$-$3.23(02)&0.091\\
normal, v $<$ 7,000&26&$-$4.17(05)&0.113&$-$3.65(03)&0.118&$-$3.56(02)&0.108&$-$3.24(02)&0.102\\
normal, v $>$ 7,000&32&$-$4.06(06)&0.104&$-$3.63(02)&0.093&$-$3.53(03)&0.091&$-$3.21(02)&0.081\\
normal, E$(B - V)_{host}\leqslant$0.15&47&$-$4.14(04)&0.126&$-$3.63(02)&0.090&$-$3.53(02)&0.087&$-$3.22(02)&0.087\\
normal, E$(B - V)_{host}\leqslant$0.06&24&$-$4.18(07)&0.177&$-$3.64(02)&0.078&$-$3.55(02)&0.074&$-$3.25(02)&0.071\\
91T/99aa-like&10&$-$4.28(08)&0.206&$-$3.70(06)&0.175&$-$3.69(04)&0.138&$-$3.35(05)&0.161\\
91bg-like&4&$-$4.14(15)&0.228&$-$3.55(08)&0.083&$-$3.56(07)&0.093&$-$3.26(07)&0.082\\
\enddata
\def\baselinestretch{1.5}
\tablenotetext{a}{N$_{BV}$ represents the number of SN Ia sample
available in the $B$ and $V$ bands.}
\end{deluxetable}

\def\baselinestretch{1.1}
\begin{deluxetable}{llcccccccccc}
\tablewidth{0pt} \tabletypesize{\tiny} \label{tab5}
\tablecaption{The absolute magnitudes of SNe Ia with Cepheid
distances.}
\tablehead{
\colhead{$SN$} & \colhead{$Galaxy$} &
\colhead{$\mu_{KP}$}&\colhead{$\mu_{KP}$(Z)}&\colhead{m$_{V}$}&\colhead{$\Delta
C_{12}$}&\colhead{E$(B -
V)_{host}$}&\colhead{m$^{0}_{V}$}&\colhead{M$^{0}_{U}$}&\colhead{M$^{0}_{B}$}
&\colhead{M$^{0}_{V}$}&\colhead{M$^{0}_{I}$}\\
 \colhead{(1)}  & \colhead{(2)}  &
 \colhead{(3)}  & \colhead{(4)}  &
 \colhead{(5)}  & \colhead{(6)}  &
 \colhead{(7)}  & \colhead{(8)}  &
 \colhead{(9)}  & \colhead{(10)} &
 \colhead{(11)} &\colhead{(12)}}
\startdata
1937C&IC 4182& 28.28(06)&28.26(06)&8.77&0.21&0.04&8.89(13)&\nodata&$-$19.38(18)& $-$19.38(16)&\nodata\\
1972E&NGC 5253&27.56(14)&27.48(14)&8.17&0.24&0.05&8.24(12)&$-$19.94(30)&$-$19.30(22)&$-$19.25(19)&\nodata\\
1974G&NGC 4414&31.10(05)&31.27(06)&12.34&0.47&0.15&11.99(22)&\nodata&$-$19.34(29)&$-$19.29(26)&\nodata\\
1981B&NGC 4536&30.80(04)&30.88(04)&11.92&0.47&0.12&11.59(08)&$-$19.79(28)&$-$19.40(21)&$-$19.29(16)&\nodata\\
1989B&NGC 3627&29.86(08)&30.04(09)&11.88&0.76&0.37&10.93(10)&$-$19.96(29)&$-$19.41(23)&$-$19.26(18)&$-$18.98(13)\\
1990N&NGC 4639&31.61(08)&31.73(08)&12.64&0.36&0.11&12.48(08)&$-$19.81(21)&$-$19.34(17)&$-$19.25(14)&$-$18.91(11)\\
1991T&NGC 4527&30.48(09)&30.58(09)&11.44&0.48&0.18&11.05(09)&$-$20.18(21)&$-$19.56(18)&$-$19.54(15)&$-$19.15(12)\\
1994ae&NGC 3370&32.22(06)&32.29(06)&12.94&0.22&0.05&13.04(08)&$-$19.80(21)&$-$19.25(16)&$-$19.23(13)&$-$18.95(09)\\
1998aq&NGC 3982&31.60(09)&31.66(09)&12.41&0.28&0.03&12.43(07)&$-$19.98(20)&$-$19.35(17)&$-$19.23(13)&$-$18.95(11)\\
1998bu&NGC 3368&29.97(06)&30.14(07)&11.80&0.64&0.36&11.03(09)&$-$19.86(20)&$-$19.17(16)&$-$19.12(13)&$-$18.94(11)\\
1999by&NGC 2841&30.74(23)&30.84(23)&13.10&1.29&0.00&11.66(11)&$-$19.71(30)&$-$19.15(27)&$-$19.15(25)&$-$18.95(25)\\
mean  &&&&&&&&$-$19.89(08)&$-$19.33(06)&$-$19.27(05)&$-$18.97(04)\\
mean &(excl.& SNe 1991T,& 99by)&&&&&$-$19.87(09)&$-$19.32(06)&$-$19.24(05)&$-$18.95(05)\\
mean &(excl.& SNe 1989B,& 98bu)&&&&&$-$19.86(10)&$-$19.34(07)&$-$19.27(07)&$-$18.94(06)\\
\enddata
  \def\baselinestretch{1.5}
\end{deluxetable}

\def\baselinestretch{1.1}
\begin{deluxetable}{cccccc}
\tablewidth{0pt} \tabletypesize{\footnotesize} \label{tab6}
\tablecaption{The Value of the Hubble constant H$_{0}$ (in km
s$^{-1}$ Mpc$^{-1}$)from SNe Ia.}\tablehead{
\colhead{Sample(N$_{h}$+N$_{c}$)$^{a}$}& \colhead{H$_{0}$(U)} &
\colhead{H$_{0}$(B)}&\colhead{H$_{0}$(V)}&\colhead{H$_{0}$(I)}&\colhead{H$_{0}$(U,B,V,I)}
}
\startdata
All(73+11)&72.1$\pm$2.9&72.8$\pm$2.1&72.1$\pm$1.8&71.8$\pm$1.4&72.1$\pm$0.9\\
normal(58+9)&71.4$\pm$3.5&72.8$\pm$2.1&72.4$\pm$2.1&71.8$\pm$1.8&72.2$\pm$1.1\\
normal, E$(B - V)_{host}$$\leq$0.15(47+7)&71.8$\pm$3.6&72.1$\pm$2.5&71.1$\pm$2.3&71.8$\pm$2.1&71.7$\pm$1.2\\
91T/99aa-like(10+1)&66.1$\pm$6.9&67.3$\pm$5.9&67.6$\pm$4.9&69.2$\pm$4.2&67.9$\pm$2.6\\
91bg-like(4+1)&76.9$\pm$12.0&75.9$\pm$9.9&76.2$\pm$9.8&72.8$\pm$8.7&75.1$\pm$5.0\\
\enddata
\tablenotetext{a}{N$_{h}$ is the number of the Hubble flow SN
sample , while N$_{c}$ is the number of the nearby calibrators.}
  \def\baselinestretch{1.5}
\end{deluxetable}

\begin{figure}
     \vspace{-1.3cm}
     \plotone{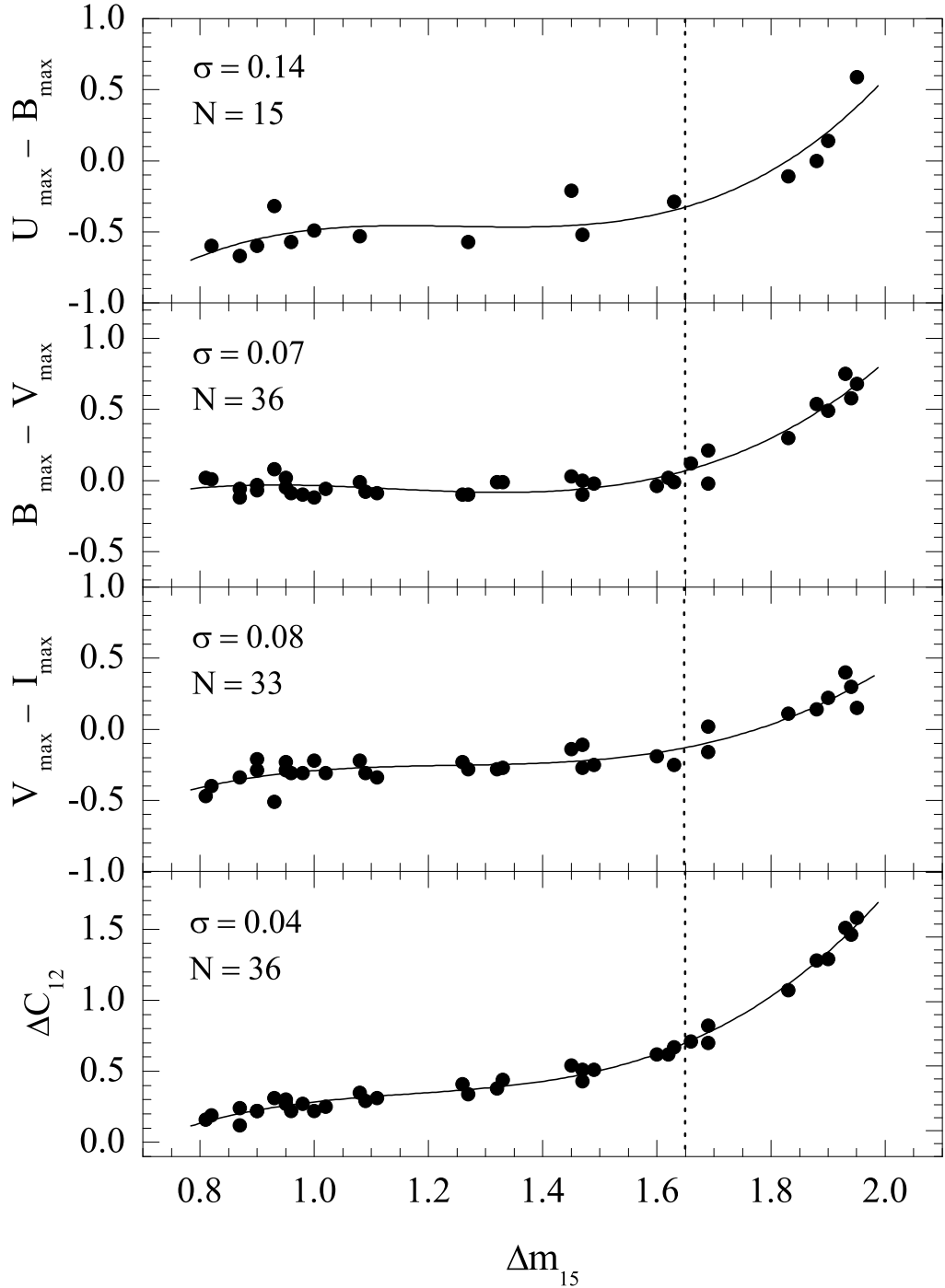}
     \vspace{-1.0cm}
     \caption{Dependence of the color parameters $U_{\hbox{\sevenrm
     max}}-B_{\hbox{\sevenrm max}}$, $B_{\hbox{\sevenrm max}}-V_{\hbox{\sevenrm max}}$,
$V_{\hbox{\sevenrm max}}-I_{\hbox{\sevenrm max}}$,, and $\Delta
C_{12}$ of SNe Ia with negligible host galaxy reddening on the
decline rate $\Delta m_{15}$.}\label{fig:one}
\end{figure}

\begin{figure}
     \vspace{-1.3cm}
     \plotone{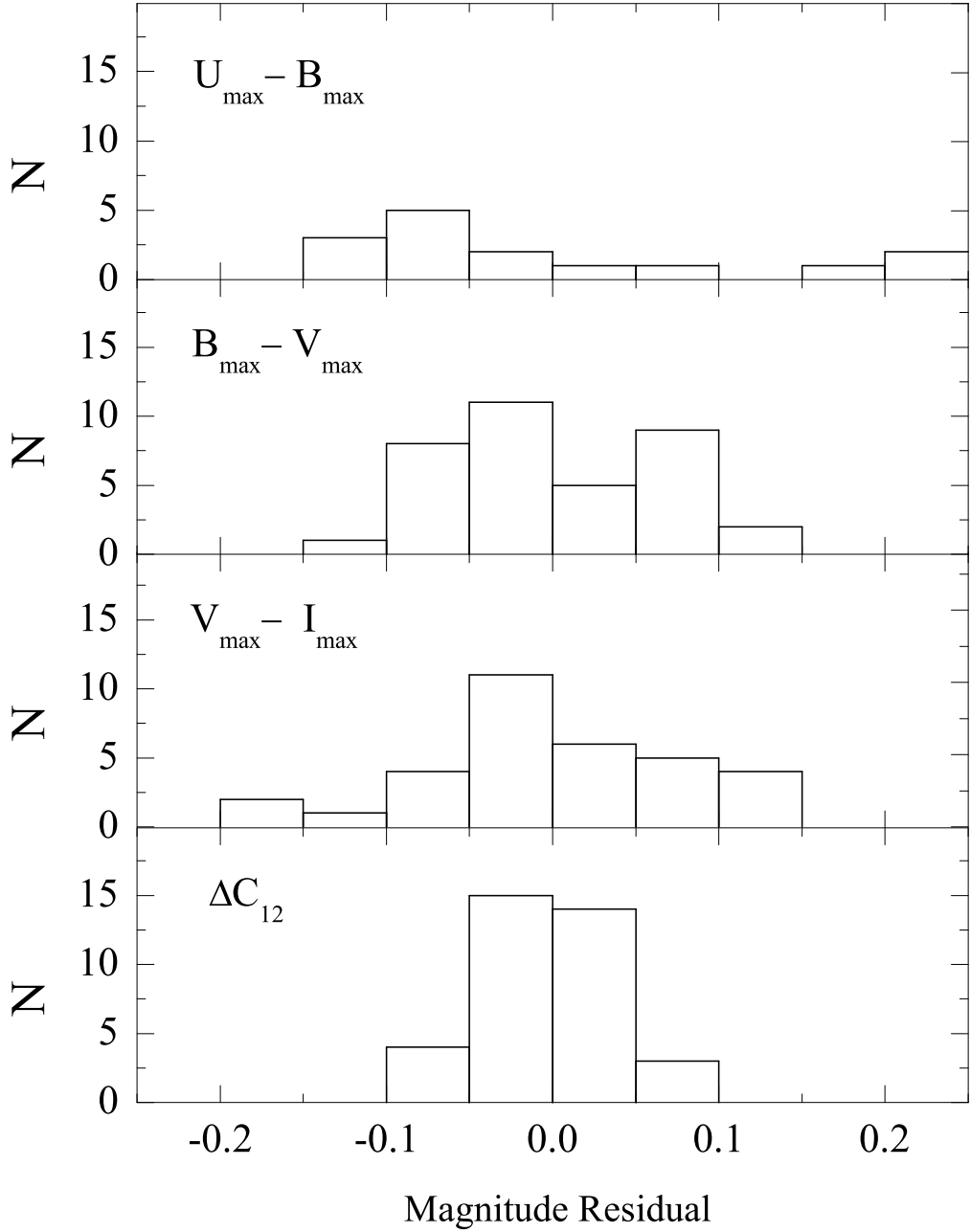}
     \vspace{-1.0cm}
     \caption{Histogram of the residual distribution for the fit to
    the color$-\Delta m_{15}$ relation as shown in Figure 1.}\label{fig:two}
\end{figure}

\begin{figure}
   \vspace{-1.0cm}
    \epsscale{.8}
   \plotone{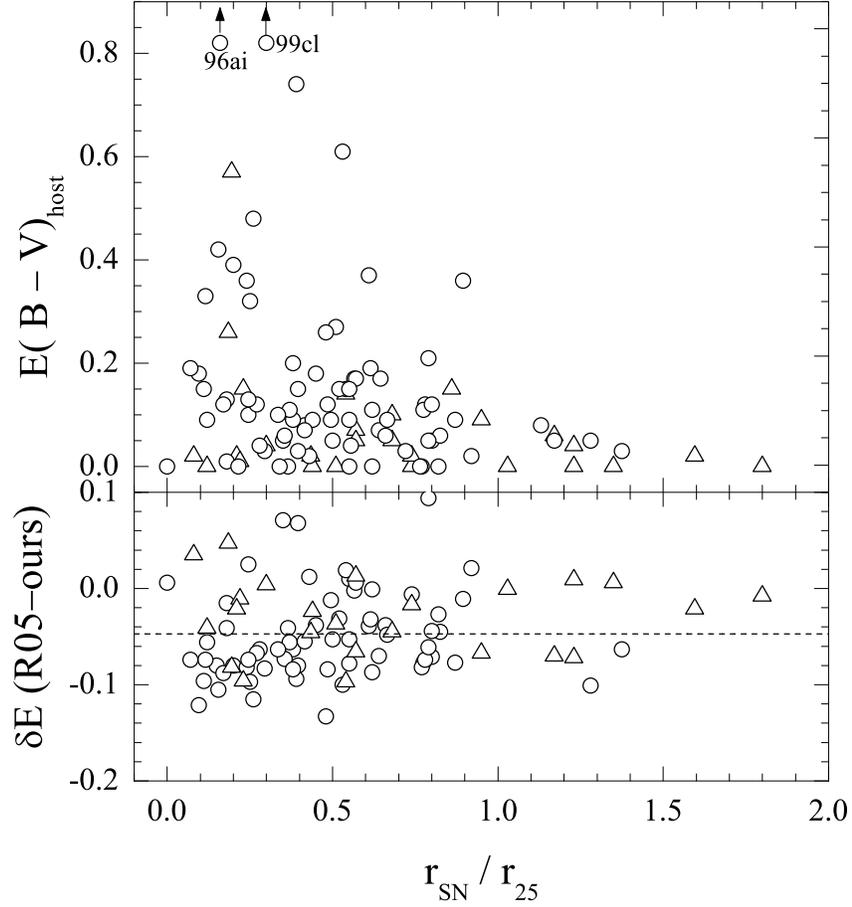}
    \epsscale{0.3}
    \vspace{-2.5cm}
    \caption{Top panel: Reddening distribution of SNe Ia in their
respective host galaxies. The circles show the SNe Ia in spiral
galaxies, and the triangles represent those in E/S0 galaxies. The
two SNe 1996ai, 1999cl as marked by the arrows have the host
galaxy reddening values of E$(B - V)_{host}$ = 1.69 and 1.20,
respectively. Bottom panel: Difference of the host galaxy
reddening values as derived between R05 and this paper. The dashed
line marks the mean value of the reddening
difference.}\label{fig:three}
\end{figure}

\begin{figure*}[t]\centering
     \vspace{-1.0cm}
     \epsscale{1.1}
     \plotone {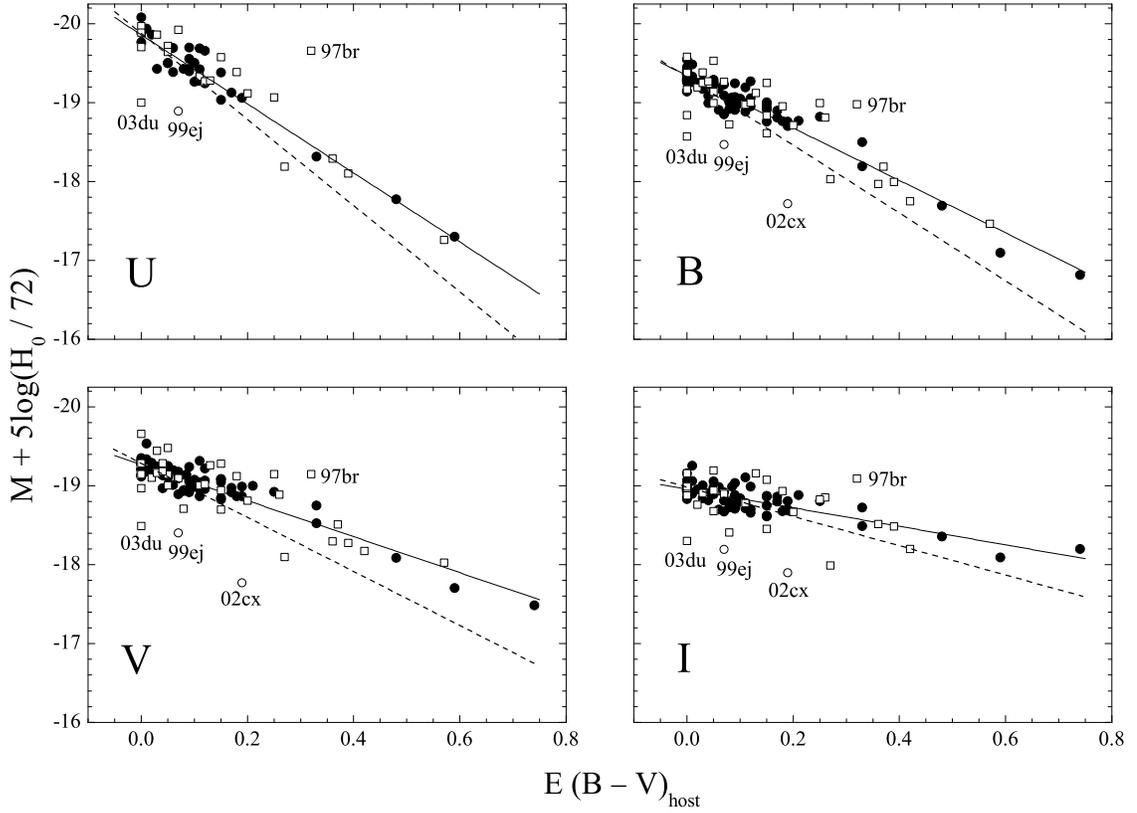}
     \vspace{-0.5cm}\caption{Dependence of the absolute magnitudes
     on the host galaxy reddening $E(B - V)_{host}$
for SNe Ia in $UBVI$ bands. The absolute magnitudes are corrected
for the Galactic absorption and the intrinsic dependence on
$\Delta C_{12}$. The circles shown here are Hubble flow SN sample.
The SNe Ia with $v\lesssim$3,000 km s$^{-1}$, represented by
squares, are not included in the fits. The two most reddened SNe
1996ai, 1999cl are not shown for the space limit. The solid lines
depict the best-fitting reddening vectors in distant galaxies,
while the dashed lines show the canonical Galactic reddening
vectors.}\label{fig:four}
\end{figure*}

\clearpage
\begin{figure*}
     \vspace{-0.3cm}
     \epsscale{.8}
     \plotone {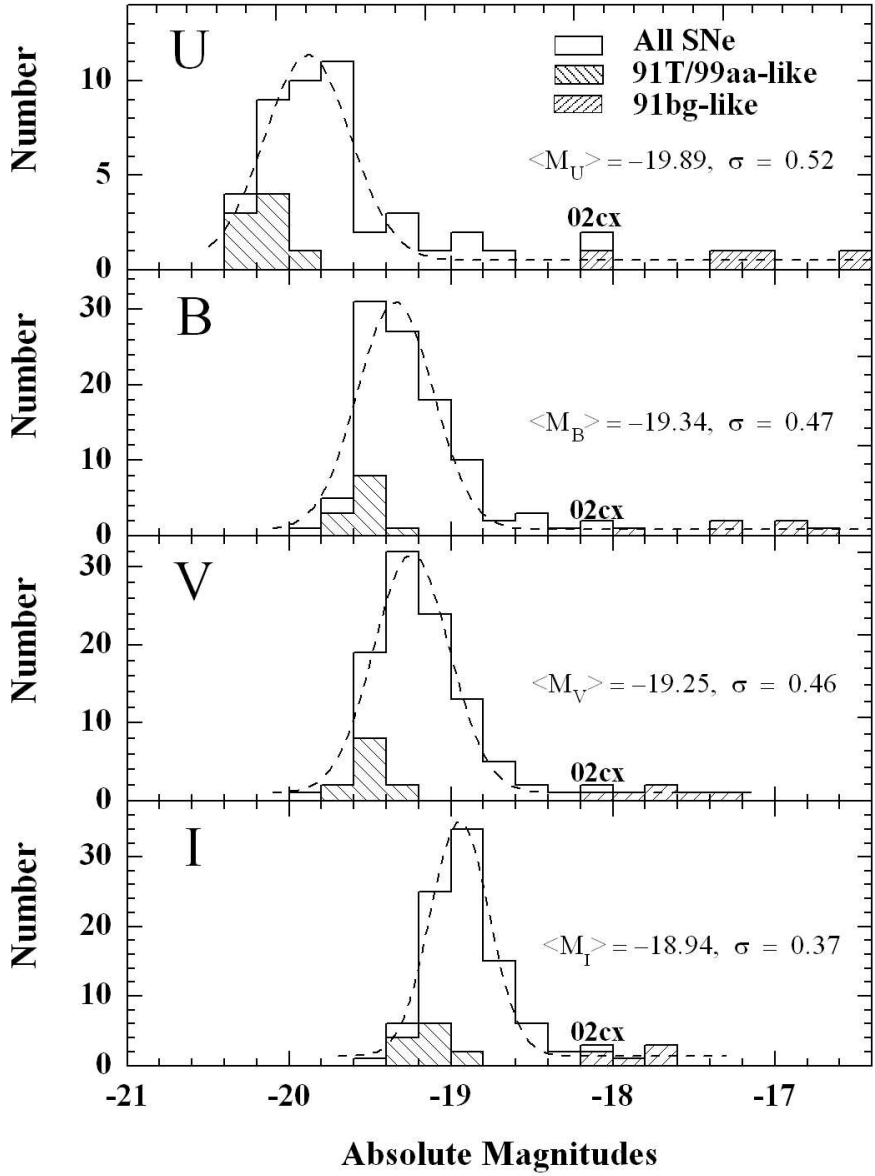}
     \vspace{0.0cm}
     \caption{Distribution of the absolute magnitudes at maximum for 109 SNe Ia in
      $UBVI$ bands. Gaussian fit (the dashed lines) to the absolute magnitudes,
      the mean values and the standard deviations are also shown.}
     \label{fig:five}
\end{figure*}


\begin{figure}
\figurenum{6} \epsscale{1.1} \plotone{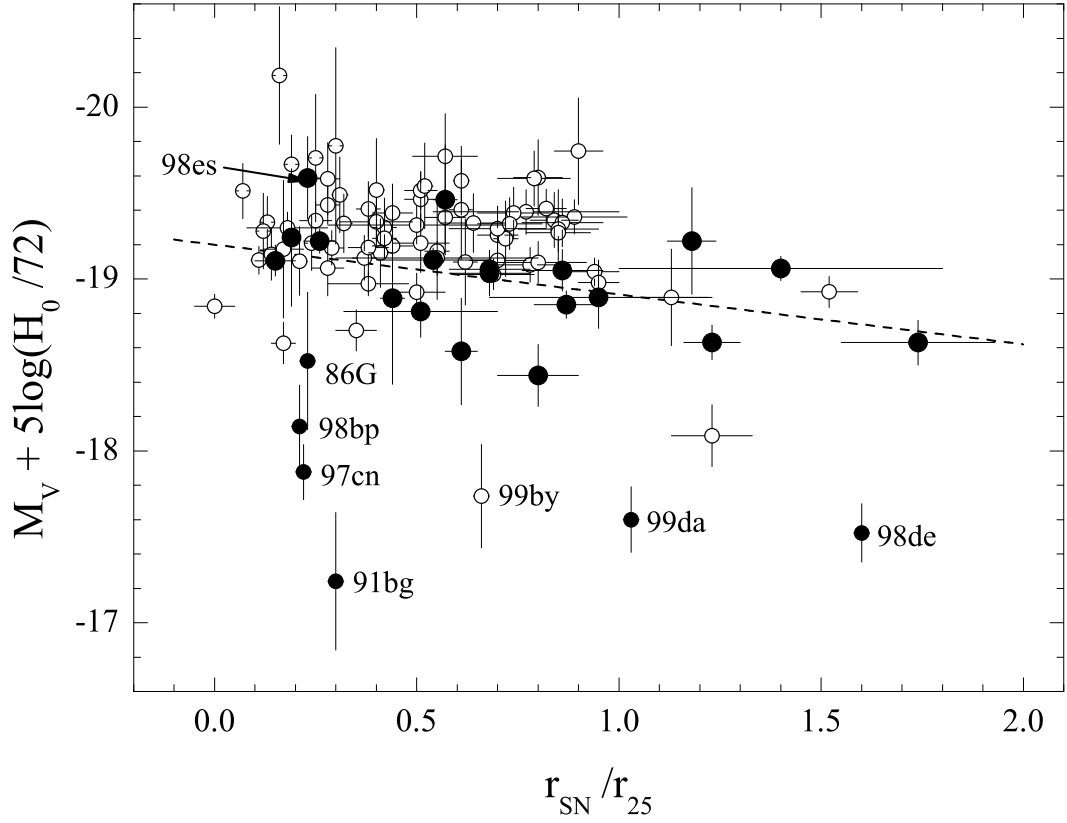} \vspace{-0.0cm}
\caption{Absolute magnitudes $M_{V}$ (fully corrected for the
extinction) plotted against the relative radial distance
r$_{SN}$/r$_{25}$. The open circles show the spirals and the
filled circles represent the E/S0 galaxies. The larger filled
circles are for the normal SNe Ia in E/S0.} \vspace{-0.0cm}
\label{fig:six}
\end{figure}

\clearpage
\begin{figure*}[t]
\begin{center}
\vspace{-1.5cm} \figurenum{7} \epsscale{1.2} \plotone{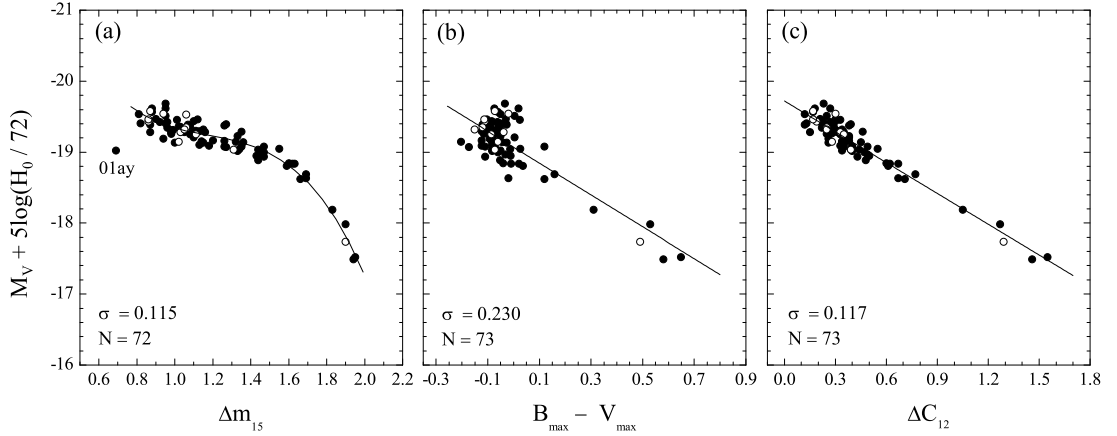}
\vspace{-6.3cm} \caption{Absolute magnitudes $M_{V}$, fully
corrected for extinction, versus the decline rate $\Delta m_{15}$,
the peak color $B-V$, and the post-maximum color $\Delta C_{12}$.
Filled circles show the Hubble flow SN sample and the open circles
represent the nearby calibrators. }\label{fig:seven}
\vspace{-0.0cm}
\end{center}
\end{figure*}

\begin{figure*}[t]
\vspace{-1.0cm} \figurenum{8} \epsscale{1.2} \plotone{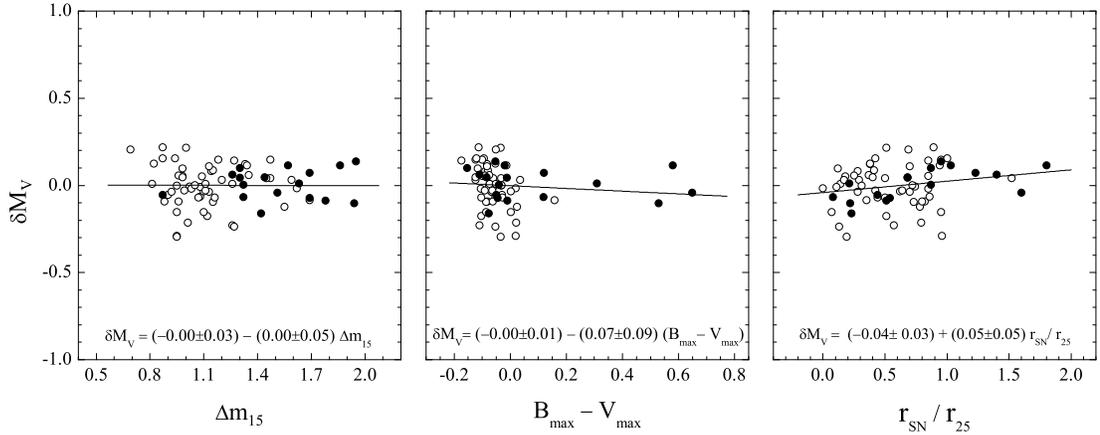}
\vspace{-6.3cm} \caption{Residuals of the $M_{V}-\Delta C_{12}$
relation fits for $M_{V}$ plotted against the decline rate $\Delta
m_{15}$, the peak color $B_{max} - V_{max}$, and the location of
the SNe Ia in their host galaxies. The open circles are for spiral
galaxies, and the filled circles are for E/S0
galaxies.}\label{fig:eight} \vspace{-0.0cm}
\end{figure*}

\clearpage
\begin{figure*}[t]
\vspace{-0.8cm} \figurenum{9} \epsscale{1.0} \plotone{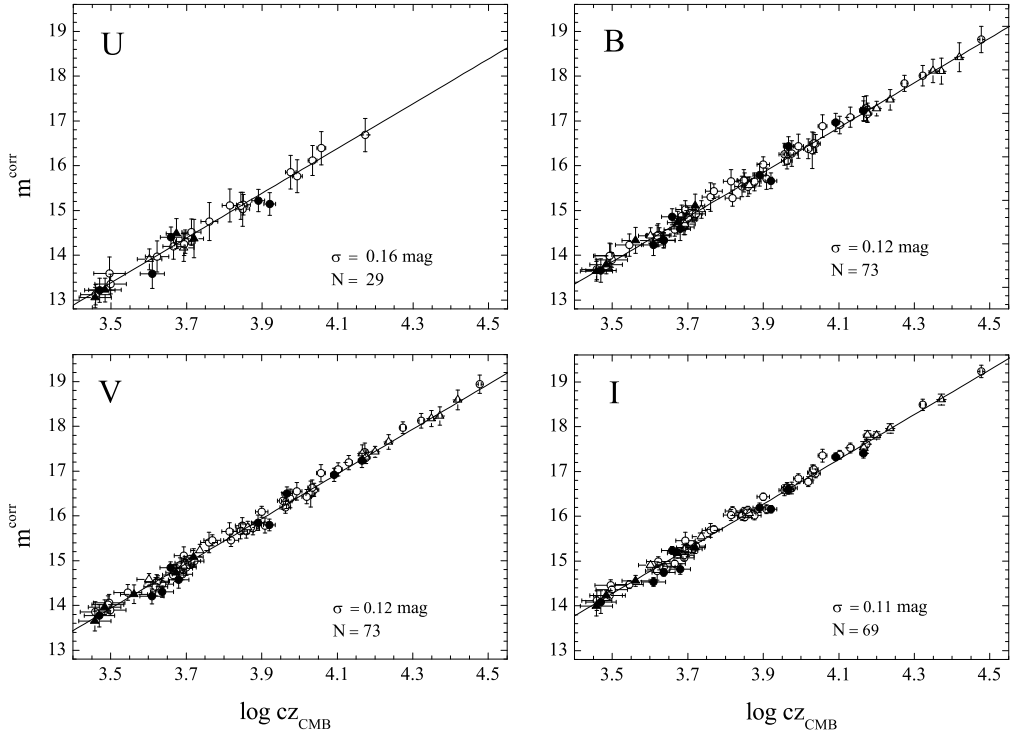}
\caption{Hubble diagrams in $U$, $B$, $V$, and $I$ for 73 SNe Ia
with 0.01$\lesssim$z$\lesssim$0.1, which were calibrated using
$\Delta C_{12}$. SNe Ia in spiral galaxies are shown with circles,
while those in E/S0 galaxies are shown with triangles. Open
symbols are for normal SNe Ia, and the filled symbols are for the
spectroscopically peculiar events. The best-fit linear regressions
are shown, with a dispersion of $\sim$0.16 mag in the $U$ band,
and $\lesssim$0.12 mag in the $BVI$ bands.}\label{fig:nine}
\vspace{-0.0cm}
\end{figure*}

\end{document}